\documentclass[dvipsnames,twocolumn]{aastex63}
\PassOptionsToPackage{usenames,dvipsnames}{xcolor}
\usepackage{tabularx}
\usepackage{amsmath}  
\usepackage{amssymb}  
\usepackage[T1]{fontenc}
\usepackage{xspace}
\usepackage{lineno}
\shorttitle{SALT3-NIR}
\shortauthors{Pierel et al.}
\defcitealias{kenworthy_salt3_2021}{K21}

\usepackage{newunicodechar,graphicx}
\DeclareRobustCommand{\okina}{%
  \raisebox{\dimexpr\fontcharht\font`A-\height}{%
    \scalebox{0.8}{`}%
  }%
}
\newunicodechar{ʻ}{\okina}

\newif{\ifchangetext}
\changetextfalse

\InputIfFileExists{\jobname-options}

\ifchangetext
  
  \newcommand{\changenote}[1]{\textcolor{blue}{ \bf #1}}x
\else
  
  \newcommand{\changenote}[1]{}
\fi

\hypersetup{
    colorlinks=True,
    citecolor=black,
    linkcolor=black,
    filecolor=magenta,      
    urlcolor=cyan,
}

\newcommand{\angstrom}{\mbox{\normalfont\AA}\xspace}

\newcommand{\SNIa}{SN\,Ia\xspace}
\newcommand{\SNeIa}{SNe\,Ia\xspace}

\def\xone{\ensuremath{x_{1}}}
\def\C{\ensuremath{\mathcal{C}}}

\newcommand{\hst}{\textit{HST}\xspace}
\newcommand{\hubble}{\textit{Hubble Space Telescope}\xspace}

\renewcommand{\roman}{\textit{Roman}\xspace}

\newcommand{\saltshaker}{\texttt{SALTshaker}\xspace}
\newcommand{\maxmodelwave}{2}
\newcommand{\numirsne}{166\xspace}

\begin{document}

\normalsize

\title{\vspace{-34pt} SALT3-NIR: Taking the Open-Source Type Ia Supernova Model to Longer Wavelengths for Next-Generation Cosmological Measurements\vspace{-5pt}}

\author[0000-0002-2361-7201
]{J.~D.~R.~Pierel}
\email{jpierel@stsci.edu}
\correspondingauthor{J.~D.~R.~Pierel} 
\affil{Space Telescope Science Institute, Baltimore, MD 21218, USA}

\author[0000-0002-6230-0151]{D.~O.~Jones}
\affil{Department of Astronomy and Astrophysics, University of California, Santa Cruz, CA 95064, USA}
\affil{NASA Einstein Fellow}

\author[0000-0002-5153-5983]{W.~D.~Kenworthy}
\affil{Department of Physics and Astronomy, The Johns Hopkins University, Baltimore, MD 21218, USA}

\author[0000-0002-5995-9692]{M.~Dai}
\affil{Department of Physics and Astronomy, The Johns Hopkins University, Baltimore, MD 21218, USA}

\author[0000-0003-3221-0419]{R.~Kessler}
\affil{Kavli Institute for Cosmological Physics, University of Chicago, Chicago, IL 60637, USA} 
\affil{Department of Astronomy and Astrophysics, University of Chicago, 5640 South Ellis Avenue, Chicago, IL 60637, USA}

\author[0000-0002-5221-7557]{C.~Ashall}
\affil{Institute for Astronomy, University of Hawai‘i, 2680 Woodlawn Dr., Honolulu, HI 96822, USA}
\affil{Department of Physics, Virginia Tech, Blacksburg, VA 24061, USA}

\author[0000-0003-3429-7845]{A.~Do}
\affil{Institute for Astronomy, University of Hawai‘i, 2680 Woodlawn Dr., Honolulu, HI 96822, USA}

\author[0000-0001-8596-4746]{E.~R.~Peterson}
\affil{Department of Physics, Duke University, Durham, NC 27708, USA}

\author[0000-0003-4631-1149]{B.~J.~Shappee}
\affil{Institute for Astronomy, University of Hawai‘i, 2680 Woodlawn Dr., Honolulu, HI 96822, USA}

\author[0000-0003-2445-3891]{M.~R.~Siebert}
\affil{Department of Astronomy and Astrophysics, University of California, Santa Cruz, CA 95064, USA}

\author[0000-0002-4843-345X]{T. Barna}
\affil{School of Physics and Astronomy, University of Minnesota, 116
Church Street SE, Minneapolis, MN 55455, USA}
\affil{Department of Physics \& Astronomy, Rutgers, State University of New Jersey, 136 Frelinghuysen Road, Piscataway, NJ 08854, USA}

\author[0000-0001-5955-2502]{T.~G.~Brink}
\affil{Department of Astronomy, University of California, Berkeley, CA 94720-3411, USA}

\author[0000-0003-0035-6659]{J.~Burke}
\affil{Las Cumbres Observatory, 6740 Cortona Dr. Suite 102, Goleta, CA 93117}
\affil{University of California, Santa Barbara, Santa Barbara, CA 93101, USA}

\author[0000-0002-0882-7702]{A.~Calamida}
\affil{Space Telescope Science Institute, Baltimore, MD 21218, USA}

\author[0000-0002-9830-3880]{Y.~Camacho-Neves}
\affil{Department of Physics \& Astronomy, Rutgers, State University of New Jersey, 136 Frelinghuysen Road, Piscataway, NJ 08854, USA}

\author[0000-0001-6069-1139]{T. de Jaeger}
\affil{Institute for Astronomy, University of Hawai‘i, 2680 Woodlawn Dr., Honolulu, HI 96822, USA}

\author[0000-0003-3460-0103]{A. V. Filippenko}
\affil{Department of Astronomy, University of California, Berkeley, CA 94720-3411, USA}

\author[0000-0002-2445-5275]{R.~J.~Foley}
\affil{Department of Astronomy and Astrophysics, University of California, Santa Cruz, CA 95064, USA}

\author[0000-0002-1296-6887]{L.~Galbany}
\affil{Institute of Space Sciences (ICE, CSIC), Campus UAB, Carrer de Can Magrans, s/n, E-08193 Barcelona, Spain}
\affil{Institut d’Estudis Espacials de Catalunya (IEEC), E-08034 Barcelona, Spain}

\author[0000-0003-2238-1572]{O.~D.~Fox}
\affil{Space Telescope Science Institute, Baltimore, MD 21218, USA}

\author[0000-0001-6395-6702]{S.~Gomez}
\affil{Space Telescope Science Institute, Baltimore, MD 21218, USA}

\author[0000-0002-1125-9187
]{D.~Hiramatsu}
\affil{Center for Astrophysics \textbar{} Harvard \& Smithsonian, 60 Garden Street, Cambridge, MA 02138-1516, USA} \affil{The NSF AI Institute for Artificial Intelligence and Fundamental Interactions}
\affil{Las Cumbres Observatory, 6740 Cortona Drive, Suite 102, Goleta, CA 93117-5575, USA} \affil{Department of Physics, University of California, Santa Barbara, CA 93106-9530, USA}

\author[0000-0002-0476-4206]{R.~Hounsell}
\affil{University of Maryland, Baltimore County, Baltimore, MD 21250, USA}
\affil{NASA Goddard Space Flight Center, Greenbelt, MD 20771, USA}

\author[0000-0003-4253-656X]{D.~A.~Howell}
\affil{Las Cumbres Observatory, 6740 Cortona Dr. Suite 102, Goleta, CA 93117}
\affil{University of California, Santa Barbara, Santa Barbara, CA 93101, USA}

\author[0000-0001-8738-6011]{S.~W.~Jha}
\affil{Department of Physics \& Astronomy, Rutgers, State University of New Jersey, 136 Frelinghuysen Road, Piscataway, NJ 08854, USA}

\author[0000-0003-3108-1328]{L.~A.~Kwok}
\affil{Department of Physics \& Astronomy, Rutgers, State University of New Jersey, 136 Frelinghuysen Road, Piscataway, NJ 08854, USA}

\author[0000-0002-2807-6459]{I.~ P\'{e}rez-Fournon}
\affil{Instituto de Astrof\'\i sica de Canarias, C/V\'\i a L\'actea, s/n, E-38205 San Crist\'obal de La Laguna, Tenerife, Spain}
\affil{Universidad de La Laguna, Dpto. Astrof\'\i sica, E-38206 San Crist\'obal de La Laguna, Tenerife, Spain}

\author[0000-0002-5391-5568]{F.~Poidevin}
\affil{Instituto de Astrof\'\i sica de Canarias, C/V\'\i a L\'actea, s/n, E-38205 San Crist\'obal de La Laguna, Tenerife, Spain}
\affil{Universidad de La Laguna, Dpto. Astrof\'\i sica, E-38206 San Crist\'obal de La Laguna, Tenerife, Spain}

\author[0000-0002-4410-5387]{A.~Rest}
\affil{Space Telescope Science Institute, Baltimore, MD 21218, USA}

\author[0000-0001-5402-4647]{D.~Rubin}
\affil{Department of Physics and Astronomy, University of Hawai`i at M{\=a}noa, Honolulu, Hawai`i 96822, USA}

\author[0000-0002-4934-5849]{D.~M.~Scolnic}
\affil{Department of Physics, Duke University, Durham, NC 27708, USA}

\author[0000-0002-1114-0135]{R.~Shirley}
\affil{Astronomy Centre, Department of Physics and Astronomy, University of Southampton, Southampton, SO17 1BJ, UK}

\author[0000-0002-7756-4440]{L.~G.~Strolger}
\affil{Space Telescope Science Institute, Baltimore, MD 21218, USA}

\author[0000-0002-1481-4676]{S.~Tinyanont}
\affil{Department of Astronomy and Astrophysics, University of California, Santa Cruz, CA 95064, USA}

\author[0000-0001-5233-6989]{Q. Wang}
\affil{Department of Physics and Astronomy, The Johns Hopkins University, Baltimore, MD 21218, USA}

\begin{abstract}
\label{sec:abstract}
\noindent A large fraction of Type Ia supernova (SN\,Ia) observations over the next decade will be in the near-infrared (NIR), at wavelengths beyond the reach of the current standard light-curve model for SN\,Ia cosmology, SALT3 ($\sim 2800$--8700\,\angstrom  central filter wavelength). To harness this new \SNIa sample and reduce future light-curve standardization systematic uncertainties, we train SALT3 at NIR wavelengths (SALT3-NIR) up to 2\,$\mu$m with the open-source
model-training software \saltshaker, which can easily accommodate future observations.  Using simulated data we show that the training process constrains the NIR model to $\sim 2$--3\% across the phase range ($-20$ to $50$\,days). We find that Hubble residual (HR) scatter is smaller using the NIR alone or optical+NIR compared to optical alone, by up to $\sim 30$\% depending on filter choice (95\% confidence). There is significant correlation between NIR light-curve stretch measurements and luminosity, with stretch and color corrections often improving HR scatter by up to $\sim20\%$. For \SNIa observations expected from the \textit{Roman Space Telescope}, SALT3-NIR increases the amount of usable data in the SALT framework by $\sim 20$\% at redshift $z\lesssim0.4$ and by $\sim 50$\% at $z\lesssim0.15$. The SALT3-NIR model is part of the open-source {\tt SNCosmo} and {\tt SNANA} \SNIa cosmology packages.
\end{abstract}




\section{Introduction}
\label{sec:intro}

Future time-domain surveys will discover samples of tens to hundreds of thousands of Type Ia supernovae (SNe\,Ia), which will be used to measure cosmological parameters such as the Hubble-Lema\^itre constant (H$_0$) and the dark energy equation of state \citep[EoS; e.g.,][]{garnavich_supernova_1998,riess_observational_1998,perlmutter_measurements_1999,scolnic_complete_2018,brout_pantheon_2022}. For cosmological measurements, SN\,Ia samples require precise photometric (and often spectroscopic) observations with a relatively high cadence over a series of weeks, and a well-trained model describing the spectral evolution of SNe\,Ia over time. This model is used for brightness standardization and subsequent luminosity-distance measurements, typically by applying corrections to an observed SN\,Ia brightness based upon the fitted light-curve shape (here called ``stretch") and color \citep[e.g., the Tripp equation;][]{tripp_two-parameter_1998}. Historically, most SN\,Ia observations have been at optical wavelengths \citep[$\sim 3000$--8000\,\angstrom; e.g.,][]{hamuy_hubble_1996,riess_bvri_1999,astier_supernova_2006,jha_ubvri_2006,jha_improved_2007,holtzman_sloan_2008,hicken_improved_2009,contreras_carnegie_2010,stritzinger_carnegie_2011,hicken_cfa4_2012,jones_measuring_2017,foley_foundation_2018,abbott_first_2019}, which has led to models of SN\,Ia spectral energy distributions (SEDs) with similar wavelength coverage \citep[e.g.,][]{guy_salt:_2005,guy_salt2:_2007,guy_supernova_2010,betoule_improved_2014,mosher_cosmological_2014,saunders_snemo_2018,leget_sugar_2020,taylor_revised_2021}. 

There are indications that the remaining scatter in residuals in a Hubble diagram made with optical distance measurements \citep[``intrinsic scatter''; e.g.,][]{scolnic_complete_2018} can be reduced by up to $\sim60\%$ using rest-frame near-infrared (NIR) SN\,Ia observations \citep{elias_observation_1986,krisciunas_hubble_2004,mandel_type_2011,dhawan_measuring_2018,avelino_type_2019}. Such an improvement would require a large sample of rest-frame NIR SN\,Ia spectra and photometry, and a well-trained model with coverage at these wavelengths to be used for brightness standardization. While there have been SN\,Ia light-curve models trained at NIR wavelengths \citep[e.g.,][]{burns_carnegie_2011,burns_carnegie_2014,mandel_type_2011,mandel_hierarchical_2020}, most large-sample cosmological analyses including SNe\,Ia for the last decade \citep{guy_supernova_2010,conley_supernova_2011,betoule_improved_2014,riess_type_2018,sako_data_2018,scolnic_complete_2018,brout_first_2019,jones_foundation_2019,brout_pantheon_2022} have used the SALT2 framework for brightness standardization \citep{guy_salt2:_2007,guy_supernova_2010}. SALT2 is a principal component analysis (PCA)-like SED time-series model with continuous wavelength coverage (2000--9200\,\angstrom), and it is trained on a mix of spectra and photometry from SNe\,Ia over a wide range in redshift $z$, light-curve stretch, and color. This model is well understood and thoroughly vetted, covers a relatively large phase range, spans slightly beyond optical wavelengths, obviates the need for K-corrections, and has a diverse, open-source training sample. If reliably extended to the NIR, SALT can continue providing a robust open-source model for the future of \SNIa light-curve fitting.  

The next decade of \SNIa surveys from space, which in particular includes the \textit{Nancy Grace Roman Space Telescope} (hereafter \textit{Roman}), will observe primarily at NIR wavelengths. The natural choice of model to harness this new wealth of data is SALT owing to its consistent use in cosmological analyses \citep[for a summary of other reasons, see][hereafter K21]{kenworthy_salt3_2021}. Although SALT2 has already been extended to NIR wavelengths  for simulation studies \citep[][hereafter P18]{pierel_extending_2018}, the resulting model was not fully trained or validated and is insufficient for cosmological analyses using real data. The SALT3 model from \citetalias{kenworthy_salt3_2021} has an improved and open-source training code and sample compared with SALT2, and its wavelength coverage in the red extended from $0.92\,\mu$m to $1.1\,\mu$m (central filter wavelength 8700\,\angstrom). However,  most low-redshift ($z\lesssim0.15$) \SNIa observations expected from \textit{Roman} are 
beyond the red end of the SALT3 wavelength range and will not currently be used for distance measurements. If this gap is not addressed with a redder rest-frame model, it will be necessary to observe a low-$z$ sample of SNe\,Ia at optical wavelengths from the ground \citep[e.g.,][]{foley_foundation_2018}, and combine these with a sample of higher-$z$ SNe\,Ia whose NIR observations will be probing rest-frame optical wavelengths. 

To include the full range of observations expected from \textit{Roman} in the robust SALT framework, and to enable future cosmological inference at NIR wavelengths, a new version of the model is needed. To address this need, we perform the first robust training of the SALT3 model to $\maxmodelwave\,\mu$m using $\numirsne$ SNe\,Ia with rest-frame NIR observations, producing a model that we label as ``SALT3-NIR.'' While the data volume and calibration are not yet of the same quality as optical data, this training sample will be continuously improved in the future and the model kept up-to-date with the open-source \saltshaker\footnote{\href{https://saltshaker.readthedocs.io}{saltshaker.readthedocs.io}} framework. This work enables \roman \SNIa data to be utilized as they become available instead of waiting until there is a sufficient \SNIa sample to train a \roman-only SALT model in the NIR, and produces an improved model for accurate simulations or fitting of early \roman observations. The full training sample used here, as well as the resulting model, will be made public once this manuscript is published. 

We begin with a discussion of the SALT3 formalism in Section \ref{sec:salt3}, followed by a description of the NIR training dataset in Section \ref{sec:data} that is used to perform the model training in Section \ref{sec:salt3-nir} and validation in Section \ref{sec:validating}. Section \ref{sec:validating} also contains an analysis of improvements for NIR cosmology relative to existing models. We conclude in Section \ref{sec:conclusion}.

\section{The SALT3 Model}
\label{sec:salt3}

\begin{figure}[t!]
    \centering
    \includegraphics[trim={.5cm 1.5cm 3cm 3cm}, clip,width=\linewidth]{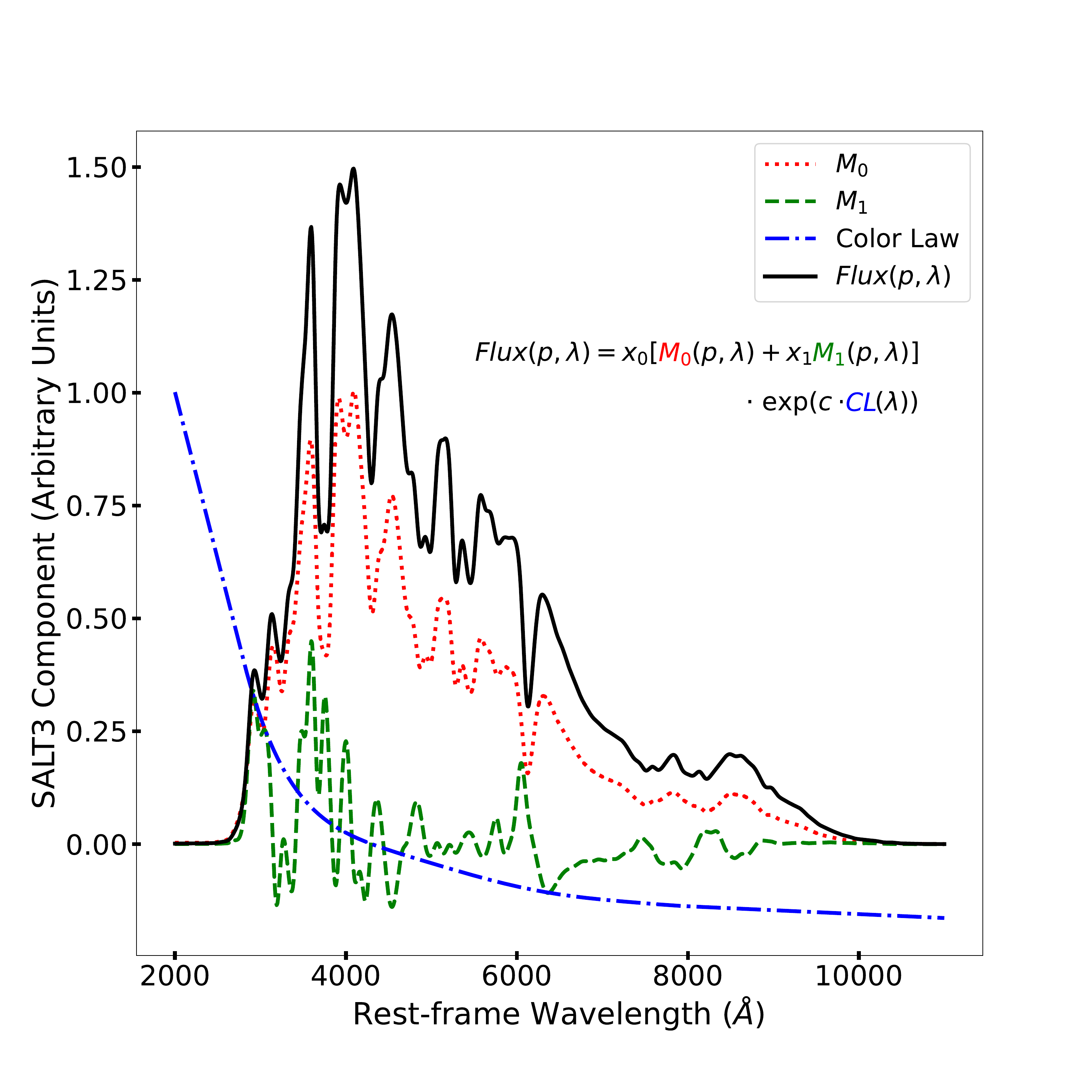}
    \caption{The primary components that make up the existing SALT3 model at peak ($p=0$) brightness (black solid). The $M_0$ component (red dotted) defines a ``baseline'' SN\,Ia spectrum, the $M_1$ component (green dashed) controls the stretch of a given SN, and the color-law (blue dash-dot) is equivalently $A_\lambda-A_B$. The terms $x_0$, $x_1$, and $c$ are free fitting parameters.}
    \label{fig:salt3}
\end{figure}

Here we briefly review the formalism of the SALT3 model; for more details, see \citetalias{kenworthy_salt3_2021}. SALT3 is based upon the original SALT framework \citep{guy_salt:_2005,guy_salt2:_2007}, which gives the spectral flux ($F$) as a function of rest-frame phase ($p$) and wavelength ($\lambda$):
\begin{align}
    \label{eq:salt}
    F(p,\lambda) = &x_0\left[M_0(p,\lambda)+x_1M_1(p,\lambda)\right] \nonumber\\&\cdot\exp{(c\cdot CL(\lambda))}\,.
\end{align}
The model is described by three components characterizing the population of SNe as determined in the training process, and three parameters characterizing each object in individual light-curve fits. The first two components are $M_0(p,\lambda)$, which represents a ``standard'' SN\,Ia, and $M_1(p,\lambda)$, which contributes a first-order linear correction to the baseline flux. Finally, there is a ``color law'' component controlling the effects of intrinsic color variation and dust, $CL(\lambda)$. $M_0$ and $M_1$ are defined as flux surfaces using an interpolated grid in phase and wavelength space that is defined by the training process, while $CL$ is defined as a polynomial (Figure \ref{fig:salt3}). The remaining terms in Equation \ref{eq:salt} are the free parameters used to fit SN\,Ia light-curve data. The parameter $x_0$ is the overall flux normalization, $x_1$ is the contribution of the $M_1$ correction term and is a measure of SN\,Ia light-curve shape or ``stretch,'' and $c$ is a measure of SN\,Ia ``color.'' These three parameters are used as part of the \citet{tripp_two-parameter_1998} equation to measure the luminosity distance to each SN\,Ia \citep[e.g.,][]{scolnic_complete_2018}. 

\subsection{SALT Distance Determinations}
\label{sub:salt_dist}
Accurately measuring \SNIa luminosity distances in the SALT framework is a two-step process. First, best-fit parameters from Equation \ref{eq:salt} ($x_0,x_1,c$) are measured for a given light curve. These parameters are used to correct the measured apparent magnitude and infer a luminosity distance using the \citet{tripp_two-parameter_1998} equation,
\begin{equation}
    \label{eq:tripp}
    \mu=-2.5\,log_{10}(x_0)+\alpha x_1-\beta c-M_0\,,
\end{equation}
where $\alpha$  $(\beta)$ is the coefficient between luminosity and stretch (color), and $M_0$ is the \SNIa absolute magnitude. In this work (Section \ref{sec:validating}), the $\alpha,\beta,M_0$ nuisance parameters are determined by the \texttt{SALT2mu} method described by \citet{marriner_more_2011}. \texttt{SALT2mu} determines the distance moduli separately from the cosmology fit by minimizing the Hubble residuals in redshift bins. This cosmology-independent method allows us to directly compare distances measured by different SALT models without considering the impact of different cosmological assumptions. When comparing SALT3 and SALT3-NIR derived distances in Section \ref{sec:validating}, we are only concerned with relative differences between the models, and therefore we ignore corrections for host-galaxy mass \citep[e.g.,][]{hamuy_hubble_1996,ivanov_relation_2000,kelly_hubble_2010,sullivan_dependence_2010} and selection bias \citep{kessler_correcting_2017}. These additional corrections are required to measure cosmological parameters \citep[e.g.,][]{scolnic_complete_2018}, but are not needed to compare Hubble residuals with different light-curve models. 

\subsection{SALT2-Extended}
\label{sub:salt2_ext}
While the SALT models have been used for optical cosmological measurements over the last $\sim15$\,yr, there has been only one attempt to extend SALT to NIR wavelengths. P18 used the template from \citet{hsiao_k_2007}, warped to photometry from the Carnegie Supernova Project (CSP), to approximately extend the SALT2 model for use in simulations in advance of future NIR surveys \citep[e.g.,][]{hounsell_simulations_2018,rose_reference_2021}. The model (here ``SALT2-Extended''), which is not trained on spectra nor a procedure to train possible relationships between the NIR and light-curve parameters $x_1$ or $c$, is not suitable for SN\,Ia cosmological parameter measurements in the NIR. Still, SALT2-Extended serves as a useful baseline for what the components of SALT3 may look like in the NIR, and provides the only data-driven means of simulating NIR photometry with an independent (but similarly conceived) model of \SNIa SED evolution. We therefore make use of SALT2-Extended in simulations for validation purposes in Section \ref{sec:validating}. 

\section{SALT3-NIR Training Sample}
\label{sec:data}
We review the \citetalias{kenworthy_salt3_2021} compilation of training data  in Section \ref{sub:data_k21}, which was used to train the SALT3 model to $\sim1.1\,\mu$m. We build upon this sample by adding a compilation of NIR photometry and spectra presented in Section \ref{sub:data_nir}. The NIR compilation includes public data (Section \ref{sub:public}), and a series of new SNe\,Ia observed in the NIR (Sections \ref{sub:sirah}- \ref{sub:dehvils}). Using this sample of photometric and spectroscopic data, we train the SALT3 model to a maximum wavelength of \maxmodelwave\,$\mu$m in Section \ref{sec:salt3-nir}, resulting in the SALT3-NIR model. The full training sample described here will be released publicly alongside the final model.

\subsection{\citetalias{kenworthy_salt3_2021} Sample}
\label{sub:data_k21}
In \citetalias{kenworthy_salt3_2021}, the most recent SALT2 training sample from \citet[][the ``Joint Light Curve Analysis,'' JLA]{betoule_improved_2014} was augmented with light curves of an additional $\sim700$ SNe\,Ia from a combination of the second data release of the Carnegie Supernova Project \citep[CSP; ][]{contreras_carnegie_2010,stritzinger_carnegie_2011,krisciunas_carnegie_2017}, the fourth data release from the Center for Astrophysics SN\,Ia program \citep[CfA4; ][]{hicken_cfa4_2012}, as well as the Foundation Supernova Survey \citep{foley_foundation_2018,dettman_foundation_2021},  the Pan-STARRS Medium Deep Survey \citep[PS1 MDS; ][]{jones_measuring_2017,villar_superraenn_2020,hosseinzadeh_photometric_2020}, and the Dark Energy Survey \citep[DES; ][]{the_dark_energy_survey_collaboration_dark_2005,abbott_first_2019,brout_first_2019}.

The spectroscopic sample was bolstered by spectra from the \texttt{KAEPORA} database \citep{siebert_investigating_2019}, the majority of which originate from the Berkeley SN\,Ia Program (BSNIP; \citealp{silverman_berkeley_2012,stahl_berkeley_2020}). Light-curve quality cuts were applied to the \citetalias{kenworthy_salt3_2021} sample requiring (in the rest frame, relative to $B$-band time of maximum light) at least four photometric measurements within $-10$ to $+35$\,days, at least one measurement between $+5$ and $+20$\,days, measurements in two filters between $-8$ and $+20$\,days, and at least one measurement within $-10$ to $-1$\,days. All photometric data zero-points and filter wavelength shifts were recalibrated using the ``Supercal'' cross-calibration methodology of \citet{scolnic_supercal_2015}, which used the $3\pi$ sky coverage of the  PS1 photometric system to determine precise offsets between PS1 (which has been measured to $<10$\,mmag precision) and other photometric systems. 

\begin{figure}[h!]
    \centering
    \includegraphics[width=.5\textwidth]{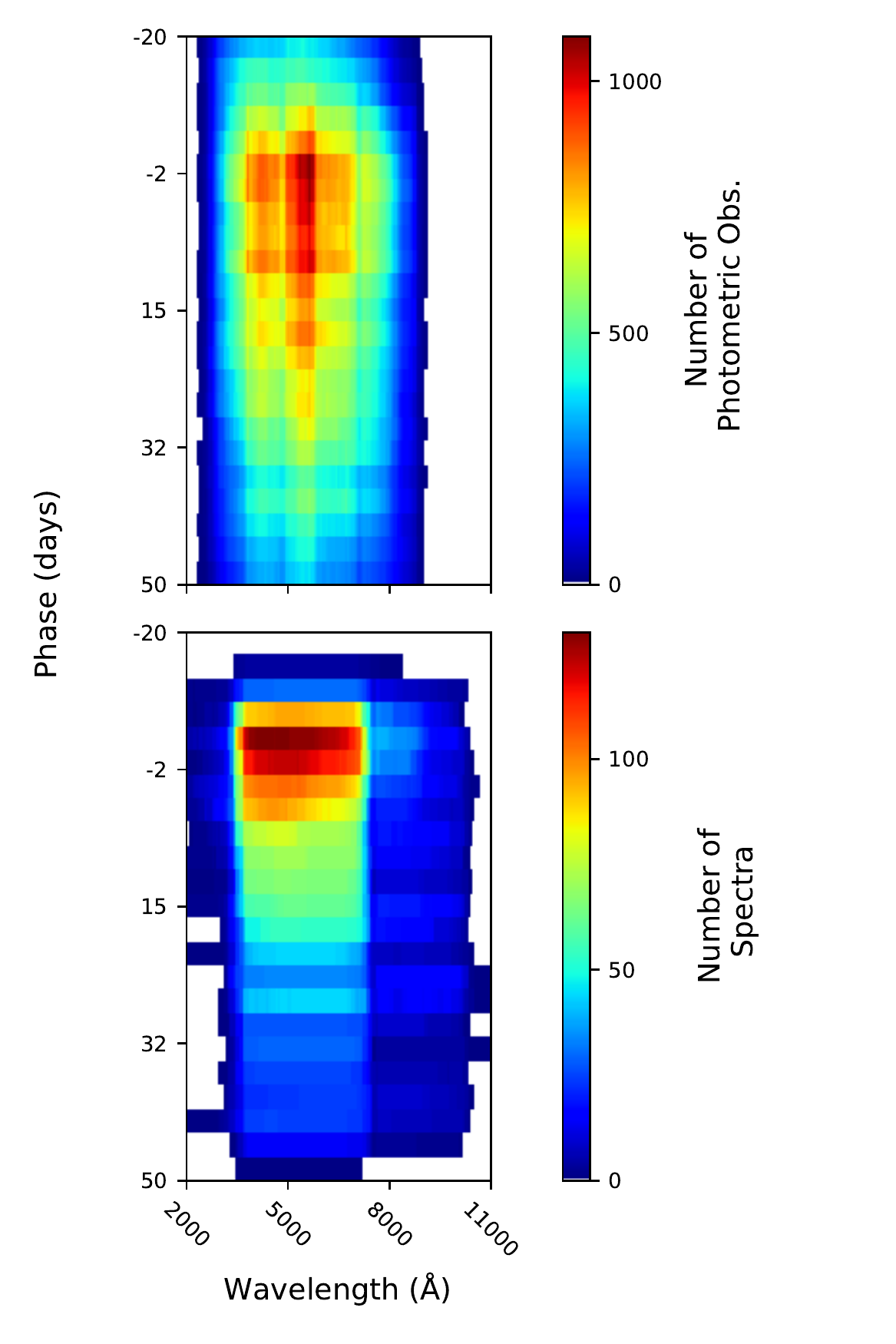}
    \caption{ The density of photometric (filter effective wavelength, top) and spectroscopic (bottom) data in the \citetalias{kenworthy_salt3_2021} sample, as a function of rest-frame wavelength and days relative to peak brightness (phase). }
    \label{fig:k21_sample}
\end{figure}

After all cuts have been applied, the  compilation of training data includes 1083 SNe\,Ia, each with a well-sampled light curve, and 1207 spectra from 380 \SNeIa. We show the density of photometric and spectroscopic data included in this compilation of optical training data in Figure \ref{fig:k21_sample}. See \citetalias{kenworthy_salt3_2021} for a full description of the composition of the sample, as well as characteristics including distributions of redshift, stretch, and color. The only additions to the \citetalias{kenworthy_salt3_2021} sample when training the optical portion of SALT3-NIR are the optical companion light curves of new SNe\,Ia with NIR observations (see next section) and updates to the photometric calibration described by \citet{brout_pantheon_2021}. This effort utilized the public Pan-STARRS stellar photometry catalog \citep{currie_evaluating_2020} to
cross-calibrate against tertiary standards released by individual \SNIa surveys, and the resulting improvements have already been applied to the SALT3 model released by \citetalias{kenworthy_salt3_2021}.\footnote{\url{https://www.github.com/sncosmo/sncosmo}}

\subsection{Near-Infrared Sample}
\label{sub:data_nir}
We have compiled a large sample of \SNeIa with publicly available data in the rest-frame NIR, which come primarily from the CfA and CSP surveys but also other SNe\,Ia reported in the literature (Section \ref{sub:public}). Additionally, we have made use of the \hubble (\hst) observations produced as part of multiple \hst programs designed to bolster space-based NIR \SNIa data (Sections \ref{sub:sirah}-\ref{sub:raisin}), and a large sample of new ground-based NIR light curves described in Section \ref{sub:dehvils}. This work does not produce any updated photometry or spectra, but instead relies upon previous or ongoing efforts to build a NIR \SNIa sample. Accordingly, we have assumed accurate calibration for all the data used here, a caveat discussed more in Section \ref{sec:conclusion}. Characteristics of the final training sample are detailed in Section \ref{sub:full_sample}.

\subsubsection{Public Data}
\label{sub:public}
The compilation of public NIR data includes 107 SNe, each with an optical+NIR light curve (CfA: \citealt{woodvasey_type_2008}; \citealt{hicken_improved_2009}; \citealt{hicken_cfa4_2012}, \citealt{friedman_cfair2_2015}; \citealt{marion_sn_2016}; CSP: \citealt{contreras_carnegie_2010}; \citealt{stritzinger_multi_2010}; \citealt{stritzinger_carnegie_2011}; \citealt{krisciunas_carnegie_2017}, and other groups: \citealt{krisciunas_optical_2003}; \citealt{krisciunas_optical_2004}; \citealt{krisciunas_type_2007}; \citealt{valentini_optical_2003}; \citealt{stanishev_sn_2007}; \citealt{pignata_optical_2008}; \citealt{leloudas_normal_2009}). This sample includes 15 NIR spectra of 12 SNe, with large gaps of coverage in rest-frame phase relative to peak that are supplemented in the following sections. 

\subsubsection{SIRAH}
\label{sub:sirah}
SIRAH\footnote{Supernovae in the Infrared Avec \textit{Hubble}:~{\it HST}-GO-15889 \& 16234} was a 2\,yr \hst program (2020--2021) that concluded in Cycle 28, having observed a new sample of SNe\,Ia in the Hubble flow ($0.02\lesssim z\lesssim0.07$) in the NIR with both \hst photometry and slitless spectroscopy. The final sample consists of 26 SNe\,Ia from a redshift range of $0.0024<z<0.07$. The design and implementation of the SIRAH program is described by Jha et al. (in prep.), and the \hst observations (and details of the surveys, photometry, and spectra) are presented by Pierel et al. (in prep.). From SIRAH, we include 485 photometric observations in 5 WFC3/IR filters, and 36 contemporaneous NIR spectra from the WFC3/IR grism ($\sim 8000$--17,000\,\AA) and Keck NIRES ($\sim 10,000$--25,000\,\AA) instruments. Ground-based optical observations of the SIRAH \SNeIa, which come from the Asteroid Terrestrial-impact Last Alert System \citep[ATLAS;][]{tonry_atlas_2018} and the Zwicky Transient Facility \citep[ZTF;][]{masci_zwicky_2019}, are also included in the sample to accurately measure the time of peak brightness, light-curve stretch, and color. 

\subsubsection{RAISIN}
\label{sub:raisin}
The RAISIN\footnote{``RAISIN'' is an anagram for ``SNIA in the IR''} program \citep{jones_cosmological_2022} was an {\it HST} program carried out during Cycles 20 and 23\footnote{{\it HST}-GO 13046 \& 14216}.  The goal of RAISIN was to observe a statistically significant sample of cosmologically useful, $z > 0.1$ SNe\,Ia in the rest-frame NIR to measure the dark energy equation-of-state parameter, $w = P/(\rho c^2)$.

RAISIN followed 46 total SNe\,Ia from the PS1 MDS and the DES SN surveys, 37 of which were used to measure $w$.  
18 SNe include three NIR epochs of data in both the $F125W$ and $F160W$ filters, while the remaining SNe include approximately three NIR epochs of data in only the $F160W$ filter. Details of the surveys and methods for deriving photometry are given by \citet{jones_cosmological_2022}. Here, as with SIRAH, we include the ground-based optical and {\it HST} NIR data for those 37 SNe in our training sample. 

\subsubsection{DEHVILS}
\label{sub:dehvils}
The DEHVILS survey (Dark Energy, H$_0$, and peculiar Velocities using Infrared Light from Supernovae) is a SN\,Ia follow-up survey on the UKIRT telescope on the summit of Maunakea, Hawai\okina i (Peterson et al., in prep.).  The survey began in March 2020, and has observed more than 100 $z < 0.1$ SNe\,Ia in the $YJH$ filters with a median of $\sim 5$ epochs of observation per SN.  SN targets are selected primarily from ATLAS \citep{tonry_atlas_2018} transient discoveries, and SIRAH targets within an accessible declination range were also prioritized.  Light curves are generated using {\tt photpipe} \citep{rest_testing_2005}, with photometric calibration tied to 2MASS \citep{skrutskie_two_2006} using the color transformations derived by \citet{hodgkin_ukirt_2009}.  DEHVILS also observed several Milky Way Cepheids in its first year, and those observations are published by \citet{konchady_h-band_2022}. The first-year SN\,Ia sample will be described by Peterson et al. (in prep.). For the purposes of being included in this training sample, we once again add the optical data available for each DEHVILS \SNIa, which primarily come from ATLAS.

\subsubsection{Full Compilation}
\label{sub:full_sample}
\begin{figure}
    \centering
    \includegraphics[width=\linewidth]{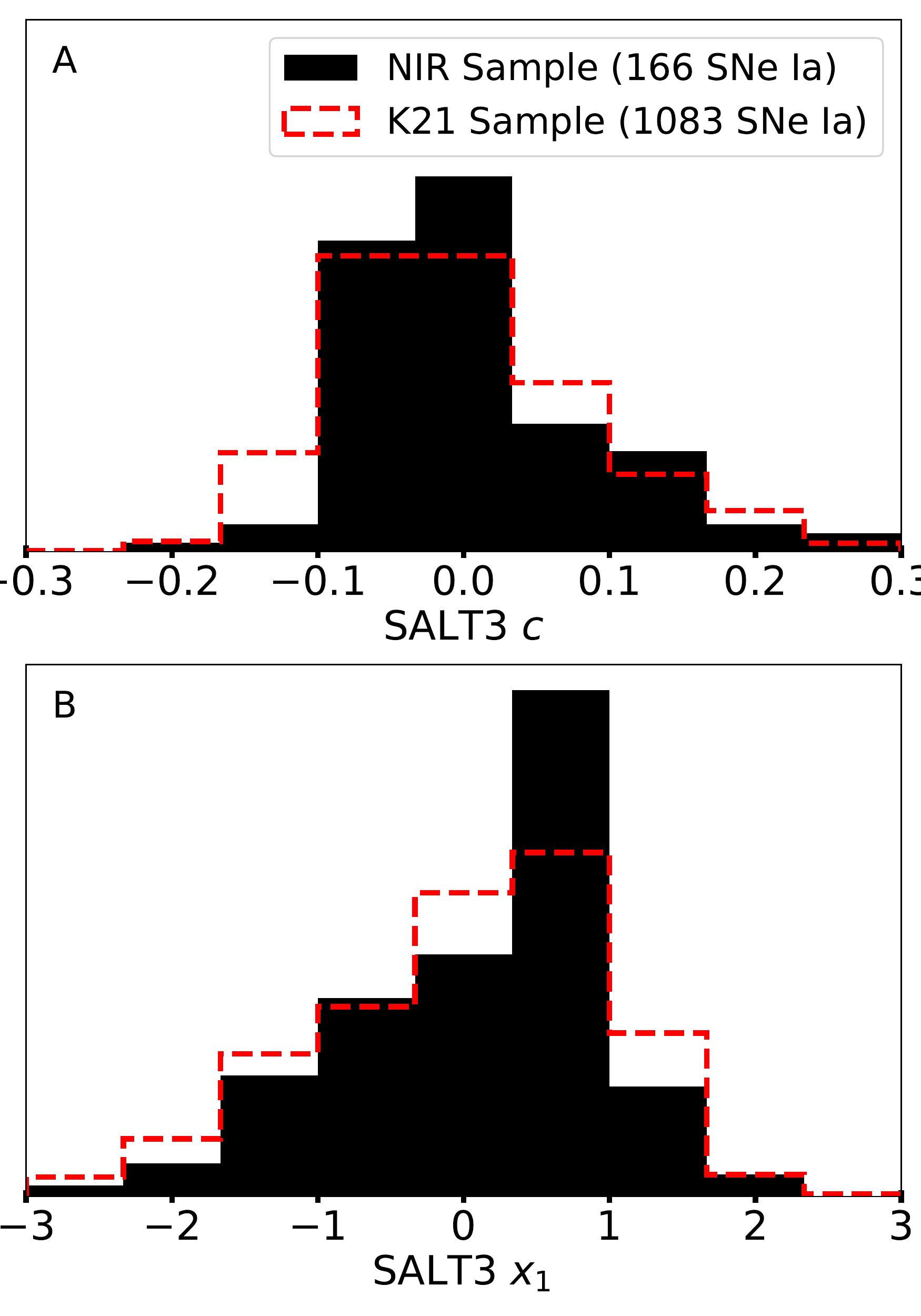}
    \caption{Distributions of SALT3 $c$ (A, top) and $x_1$ (B, bottom) for the \citetalias{kenworthy_salt3_2021} sample (red dashed line) and the added NIR sample (black). \label{fig:nir_c_x1}}
\end{figure}

\begin{figure}[h!]
    \centering
    \includegraphics[width=.5\textwidth]{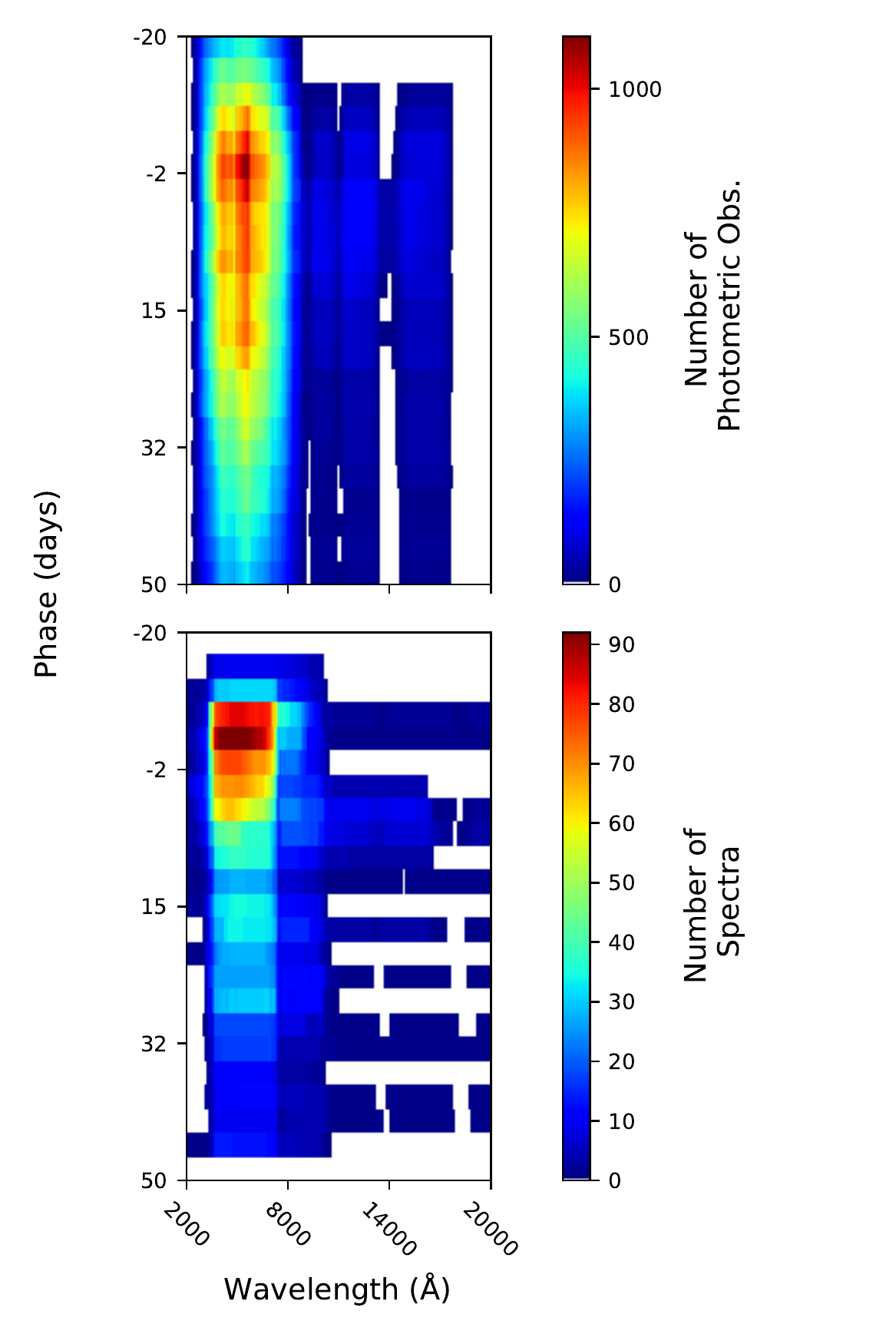}
    \caption{The density of photometric (filter effective wavelength, top) and spectroscopic (bottom) data in the \citetalias{kenworthy_salt3_2021}+NIR training sample, as a function of rest-frame wavelength and days relative to peak brightness (phase).}
    \label{fig:p22_sample}
\end{figure}

\begin{figure*}
    \centering
    \includegraphics[trim={2.2cm 1cm .5cm .25cm},clip,width=\textwidth]{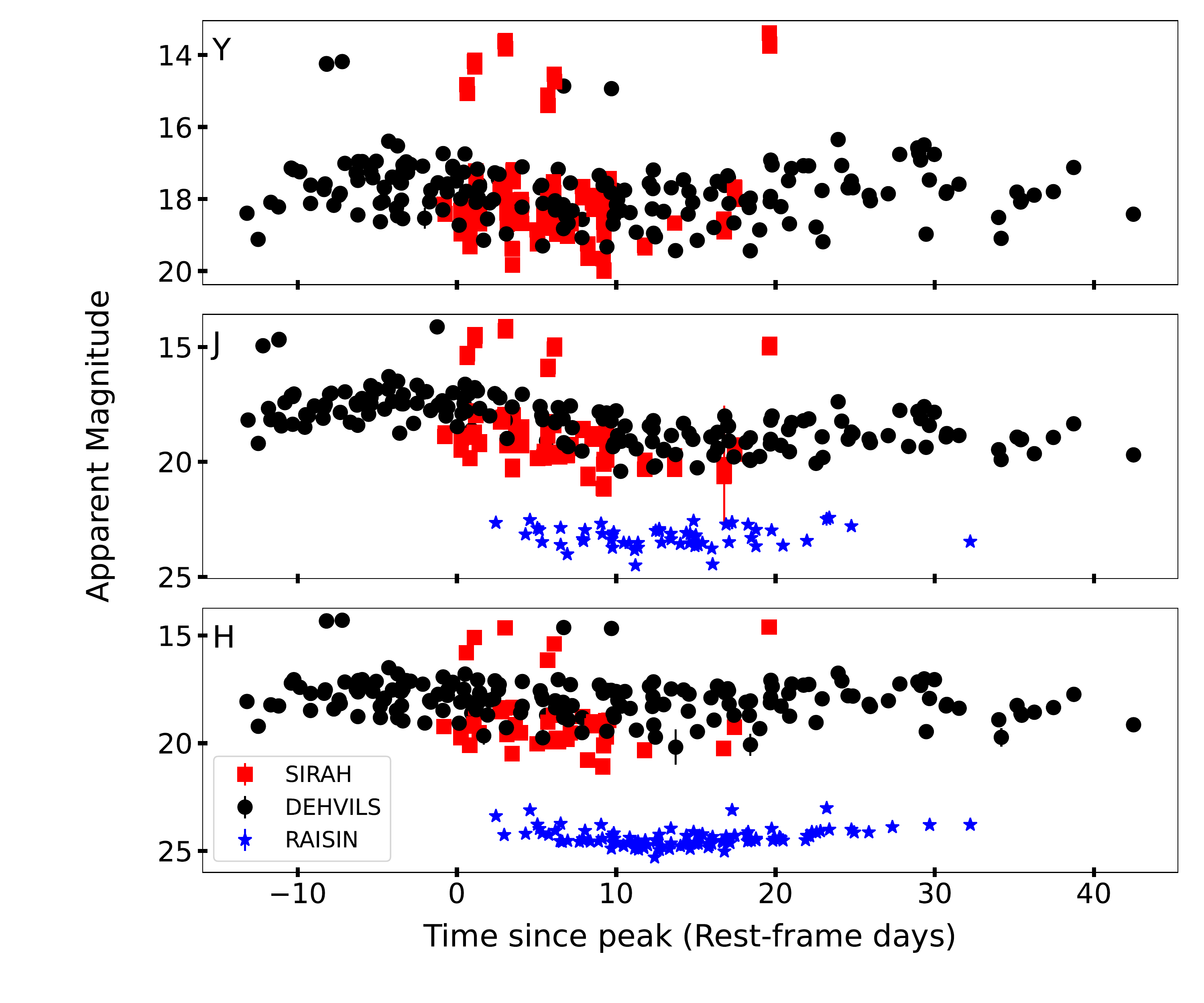}
    \caption{Additional $YJH$ data from the SIRAH, RAISIN, and DEHVILS surveys described in Sections \ref{sub:sirah}--\ref{sub:dehvils}. These light curves are combined with the legacy public NIR data listed in Table \ref{tab:p22} to create the full SALT3-NIR training sample.}
    \label{fig:nir_lcs}
\end{figure*}

Characteristics of the full NIR training sample are listed in Table \ref{tab:p22}, which describes only the \SNeIa with NIR data. The overall coverage in phase and wavelength space can be seen in Figure \ref{fig:p22_sample}. In total, we have made use of \numirsne SNe\,Ia with NIR data to train SALT3-NIR. Each of these SNe has a well-sampled optical light curve, which is used to constrain the time of maximum light, light-curve stretch, and color (Figure \ref{fig:nir_c_x1}). We have applied no additional optical cuts on the training sample, beyond those described by \citetalias{kenworthy_salt3_2021} and outlined in Section \ref{sub:data_k21}. For the NIR, we require at least two epochs of photometry in at least one NIR filter (i.e., $YJH$) within the rest-frame phase range of SALT3-NIR ($-20$ to $50$\,days). The additional photometric data coverage from the \SNeIa in Sections \ref{sub:sirah}--\ref{sub:dehvils} is shown in Figure \ref{fig:nir_lcs}, and an example of the added NIR spectra is shown in Section \ref{sub:val_sims}.

\begin{table*}
\caption{\label{tab:p22} Full details of the NIR training sample. For information about individual surveys and their references, see Section \ref{sec:data}. }
\begin{tabular*}{\textwidth}{@{\extracolsep{\stretch{1}}}*{6}{r}}
\toprule
 Survey & $N_{\rm SN}$ & $Y^a$ & $J^a$ & $H^a$ & NIR Spectra$^b$\\
\hline
CfA1&2&0&0&0&2\\
CfA2&3&0&7&3&1\\
CfA3&13&140&228&184&5\\
CfA4p1&2&42&51&58&0\\
CfA4p2&2&4&24&37&0\\
CSP&25&572&473&422&1\\
Foundation&4&0&0&0&4\\
Other&38&398&838&828&2\\
\hline
Previous Public Total&89&1,156&1,621&1,532&15\\
\hline
DEHVILS&20&183&173&175&0\\
RAISIN&34&0&57&105&0\\
SIRAH&23&173&174&83&36\\
\hline
Additional Total&77&356&404&363&36\\
\hline
\hline
\textbf{NIR Training Sample Total}&\textbf{166}&\textbf{1,512}&\textbf{2,025}&\textbf{1,895}&\textbf{51}\\
\end{tabular*}

$^a$Number of photometric observations in this rest-frame filter.\\
$^b$Includes spectra with coverage $\lambda\geq1\,\mu$m.
\end{table*}

\section{Creating SALT3-NIR with \saltshaker}
\label{sec:salt3-nir}

In an effort to standardize the methodology of training and validating SALT spectrophotometric models, \citet{kenworthy_salt3_2021} presented the \saltshaker software program. \saltshaker is a flexible and open-source training and validation pipeline written in Python, which we use here to train the new SALT3-NIR model that extends into the rest-frame NIR.  \saltshaker enables the astronomy community to retrain future
iterations of SALT3-NIR.

Here we use the same \saltshaker implementation as in \citetalias{kenworthy_salt3_2021}, with minor changes to accommodate the NIR (see following section). Briefly, the general training procedure builds an empirical SALT3-NIR model by using photometric and spectroscopic data from a SN sample to fit the model parameters with the following methods (see Section 2 of \citetalias{kenworthy_salt3_2021}).

\begin{itemize}
    \item The $M_0$ and $M_1$ principal components are constructed using a second-order B-spline basis, where the values of the knots are determined by fitting the data.
    \item The SALT3-NIR color law is determined by fitting a fourth-order polynomial with a linear extension.
    \item The SN\,Ia parameters $x_0$, $x_1$, and $c$ are fit simultaneously with the model parameters.
    \item The SN\,Ia spectra are recalibrated by using a polynomial (as a function of wavelength) that warps each spectrum to match the SED model.  This spectral warping allows the training procedure to account for systematic shifts in the spectral calibration.
    \item The SALT3-NIR error model is constructed using a zeroth-order B-spline basis, equivalent to binning the data in phase and wavelength.
    \item The SALT3-NIR color dispersion model uses a single sixth-order polynomial to model the wavelength-dependent scatter simultaneously in each independent filter.
\end{itemize}

\saltshaker uses an iterative Levenberg-Marquardt algorithm to minimize the $\chi^2$ between the model, constructed using the methods above, and the data, while estimating the free parameters of the model.  The $\chi^2$ includes contributions from the SN photometry and spectra, with additional ``regularization" terms that penalize the $\chi^2$ for high-frequency $M_0$ and $M_1$ variations as a function of phase and wavelength, and which ensures the model is separable in phase and wavelength (see \citetalias{kenworthy_salt3_2021}, Equation 11).

A number of model definitions are used to avoid degeneracies between the model components ($M_0, M_1, CL$) and model parameters ($x_0,x_1,c$). An example of such a degeneracy is an increase in the value of $x_0$ and simultaneous decrease in the amplitudes of $(M_0,M_1)$, which would leave the overall model flux unchanged.  We add priors that fix the peak $B$-band magnitude to an arbitrary value to avoid an $M_0$--$x_0$ degeneracy, and that force $(\mu_{x_1},\sigma_{x_1})=(0,1)$ to avoid an $M_1$--$x_1$ degeneracy.  Another prior forces the final $c$ parameter distribution of the training sample to be 0 ($\mu_c=0$), and we anchor the value of the color law at the central wavelengths of the Bessell $B$ and $V$ bands to avoid degeneracies between $c$ and the color law.  A final prior reduces the correlation between the $x_1$ and $c$ distributions, to avoid confusion of color differences with the $M_1$ component. Each of these priors, as well as more details of their purpose, is described in Section 2.1 of \citetalias{kenworthy_salt3_2021}. 

The \saltshaker procedure alternates between fitting the model parameters and fitting the error model, where error-model determination is performed by maximizing the log-likelihood with {\tt iMinuit}\footnote{\url{https://github.com/iminuit/iminuit}}, until the fitting converges.  Further details of the \saltshaker training process are given by \citetalias{kenworthy_salt3_2021}, with their Figure 2 giving an overview schematic of the training procedure. All SALT3 comparisons in Section \ref{sec:validating} use this model training process, with the updated calibration discussed at the end of Section \ref{sub:data_k21}\footnote{This SALT3 version is now available in SNCosmo.}.

\begin{figure*}[h!]
    \centering
    \includegraphics[width=\textwidth]{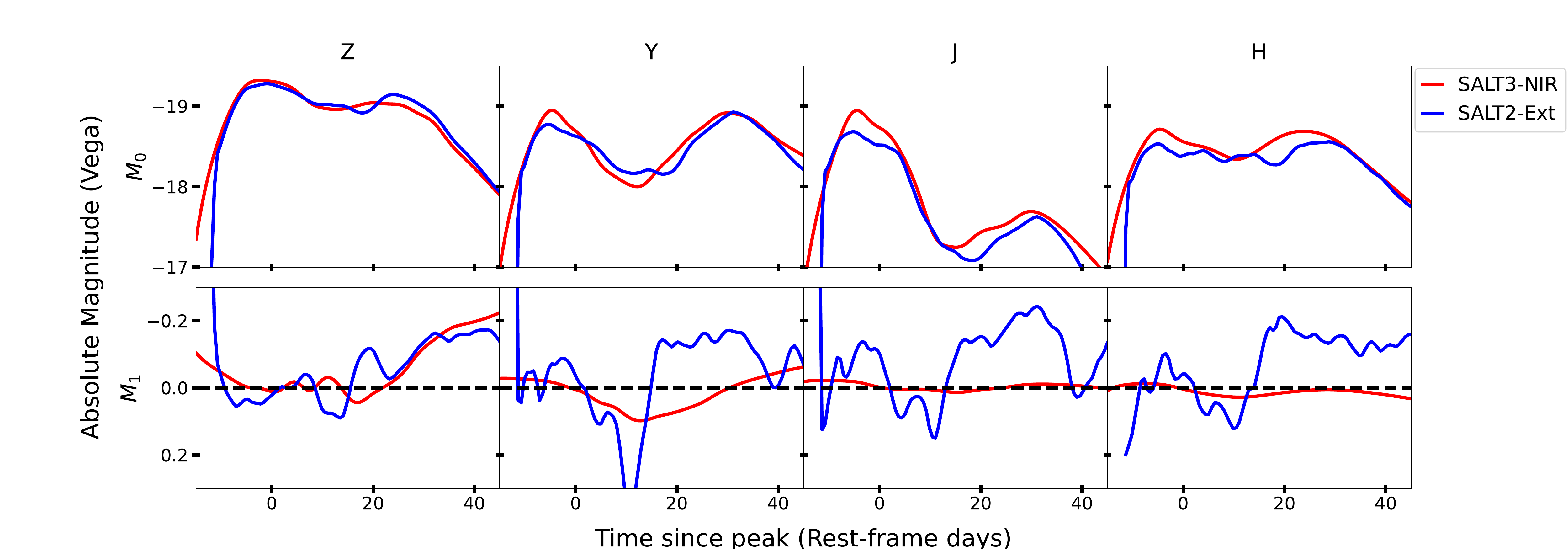}
    \caption{The $M_0$ (top row) and $M_1$ (bottom row) model components integrated over the $zYJH$ bandpasses. The new SALT3-NIR model is shown in red, while the SALT2-Extended model (previously used for SALT2 simulations in the NIR) is shown in blue. }
    \label{fig:s3nir_salt2ext}
\end{figure*}

\begin{figure*}[h!]
    \centering
    \includegraphics[width=\textwidth,trim={2.5cm .5cm 1cm 2cm},clip]{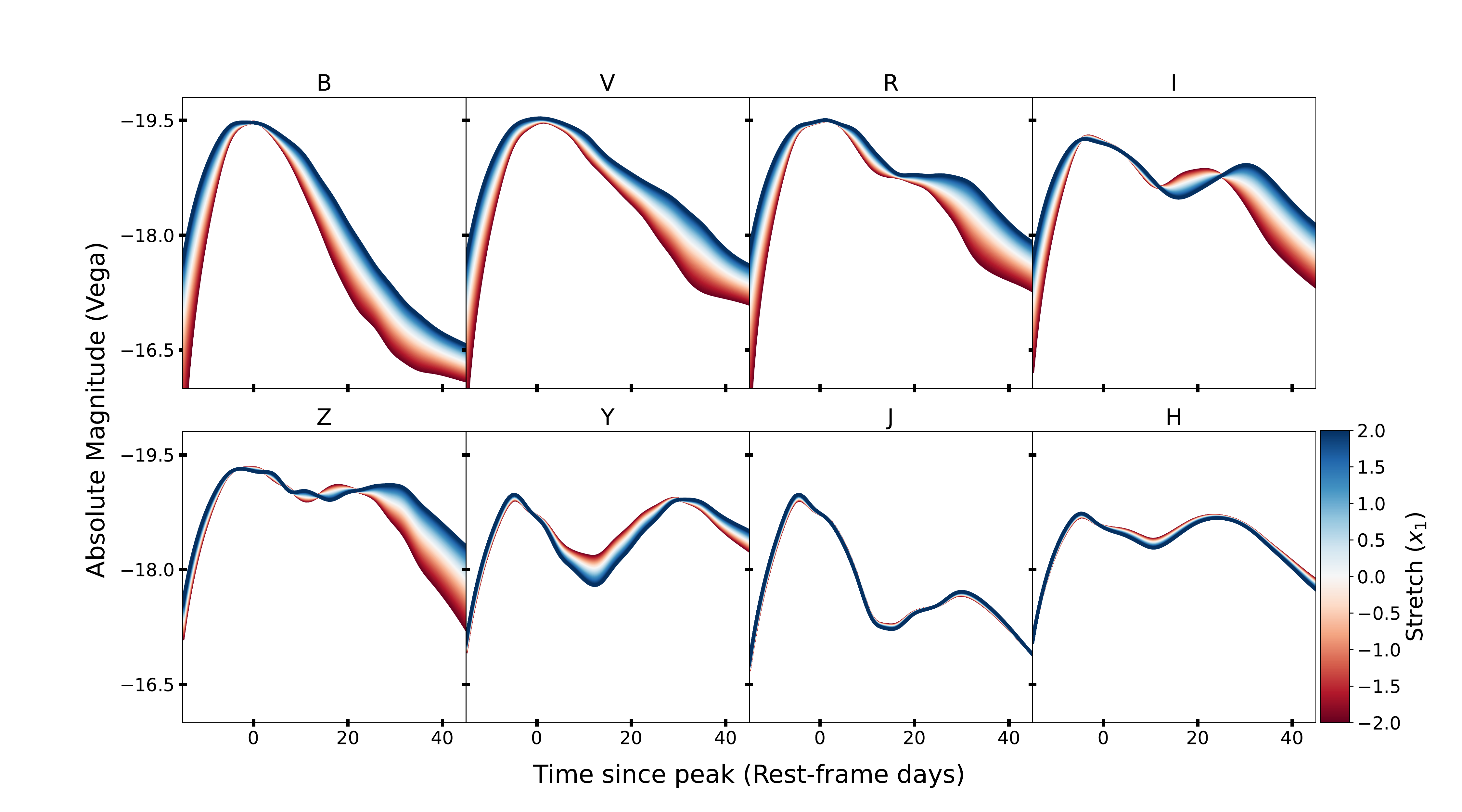}
    \caption{ The SALT3-NIR model flux integrated over optical+NIR bandpasses, as a function of phase and the light-curve stretch parameter, $x_1$. }
    \label{fig:salt3-opt-nir}
\end{figure*}

\subsection{\saltshaker in the NIR}
\label{sub:training}

Extending the wavelength range of SALT3 requires some adjustments to the \saltshaker parameters used by K21.  We first extend the $M_0$ and $M_1$ wavelength ranges to a maximum of 20,000\,\AA\ to fully encompass the red edge of all photometric filters and spectra in our sample. Next, while \citetalias{kenworthy_salt3_2021} fit the color law polynomial up to 8000\,\AA\ and allowed only linear variation in the range 8000--11,000\,\AA,
we use the additional constraining power of NIR data to fit the color-law polynomial up to
12,500\,\AA. This corresponds to roughly the central $J$-band filter wavelength, and we similarly allow only linear variation in the range 12,500--20,000\,\AA.  

\subsection{SALT3-NIR}
\label{sub:trained_salt3nir}
\begin{figure}[h!]
    \centering
    \includegraphics[width=\linewidth]{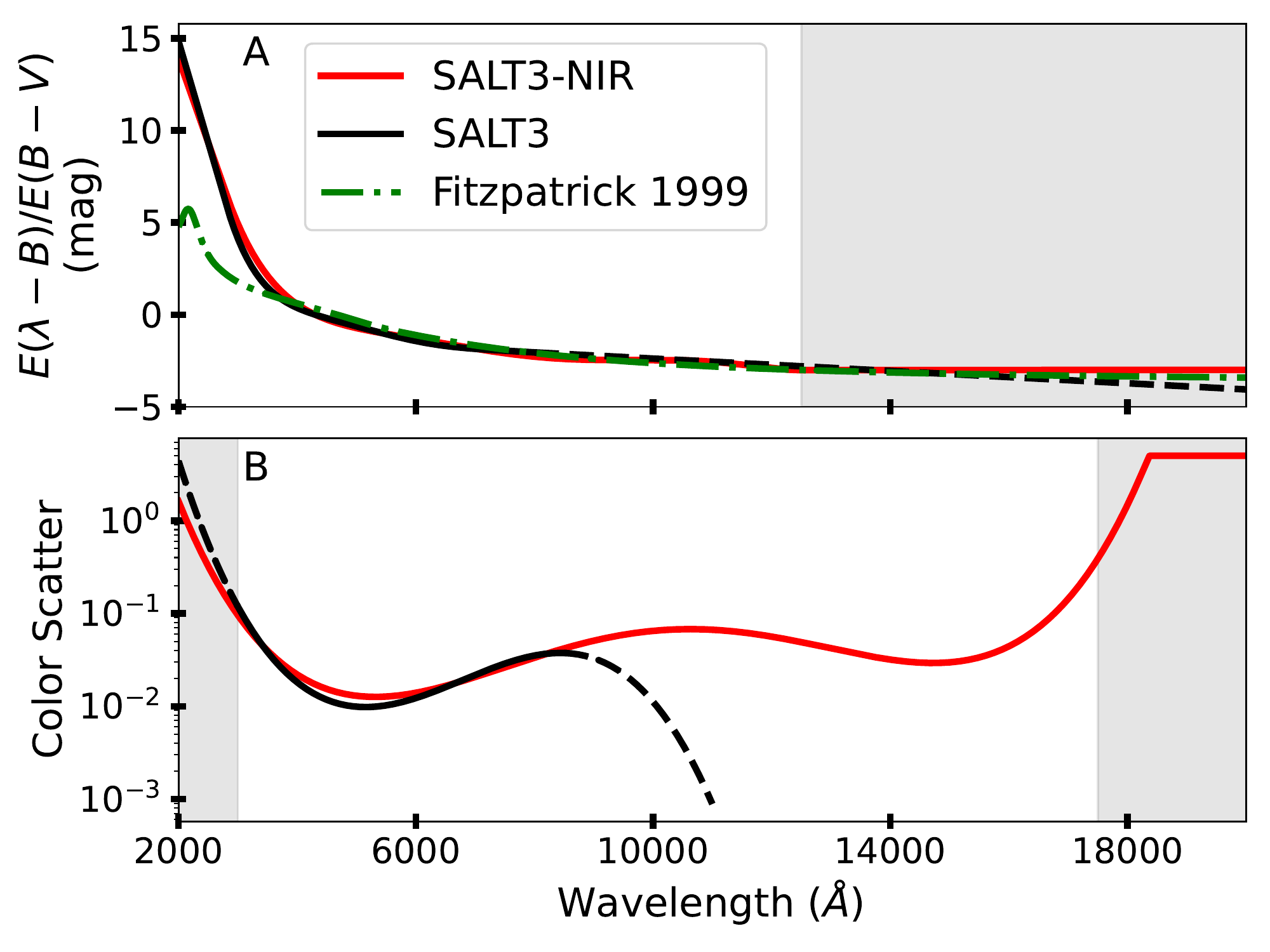}
    
    \caption{Color law (top) and color scatter (bottom) for SALT3 (black) and SALT3-NIR (red). The dashed region of the SALT3 curve corresponds to the linear regime (color law) or no constraint (color scatter). The same regions are depicted by gray shading for SALT3-NIR. The dust law from \citet{fitzpatrick_correcting_1999} (green dash-dot, $R_V=3.1$) is shown in the upper panel for comparison.}

    \label{fig:salt3-nir_errs}
\end{figure}

We train SALT3-NIR using \saltshaker in the manner described in the preceding sections, and with the full training sample from Section \ref{sub:full_sample}. Features of the resulting model are shown in Figures \ref{fig:s3nir_salt2ext}--\ref{fig:salt3-nir_errs}. In particular, Figure \ref{fig:s3nir_salt2ext} compares SALT3-NIR to SALT2-Extended (Section \ref{sub:salt2_ext}). Significant differences are obvious in the smoothness of the model $M_1$ component (no regularization was attempted for SALT2-Extended), and while generally the agreement between the two models is quite good, there are significant differences in the $JH$-band peak magnitudes. The impact of these differences on the reference survey simulations for the \textit{Roman}  \citep{hounsell_simulations_2018,rose_reference_2021} is currently being investigated (Macias et al., in prep.).

Figure \ref{fig:salt3-opt-nir} shows the model flux integrated over each bandpass for the full optical+NIR range SALT3-NIR, as a function of the light-curve stretch parameter, $x_1$. This relation makes the potential of NIR cosmology apparent, as the $JH$ bands have extremely uniform absolute magnitudes and only a weak correlation with light-curve stretch compared to the optical. The sharp variation in the $z$ band around $+20$\,days is likely a training artifact, perhaps due to a lack of sufficient coverage in the $x_1$ parameter at this phase/wavelength combination. Figure \ref{fig:salt3-nir_errs} compares the SALT3 and SALT3-NIR color law and color scatter, which describes an intrinsic variance of the \SNIa population unexplained by the model, with significant advancements in the NIR. As we still train the optical portion of the model simultaneously with the NIR, there are differences at bluer wavelengths between SALT3 and SALT3-NIR; these are discussed in Section \ref{sec:conclusion}.

\section{Validating SALT3-NIR}
\label{sec:validating}

\begin{figure}
    \centering
    \includegraphics[trim={0cm 4cm 1.5cm 5cm},clip,width=.87\linewidth]{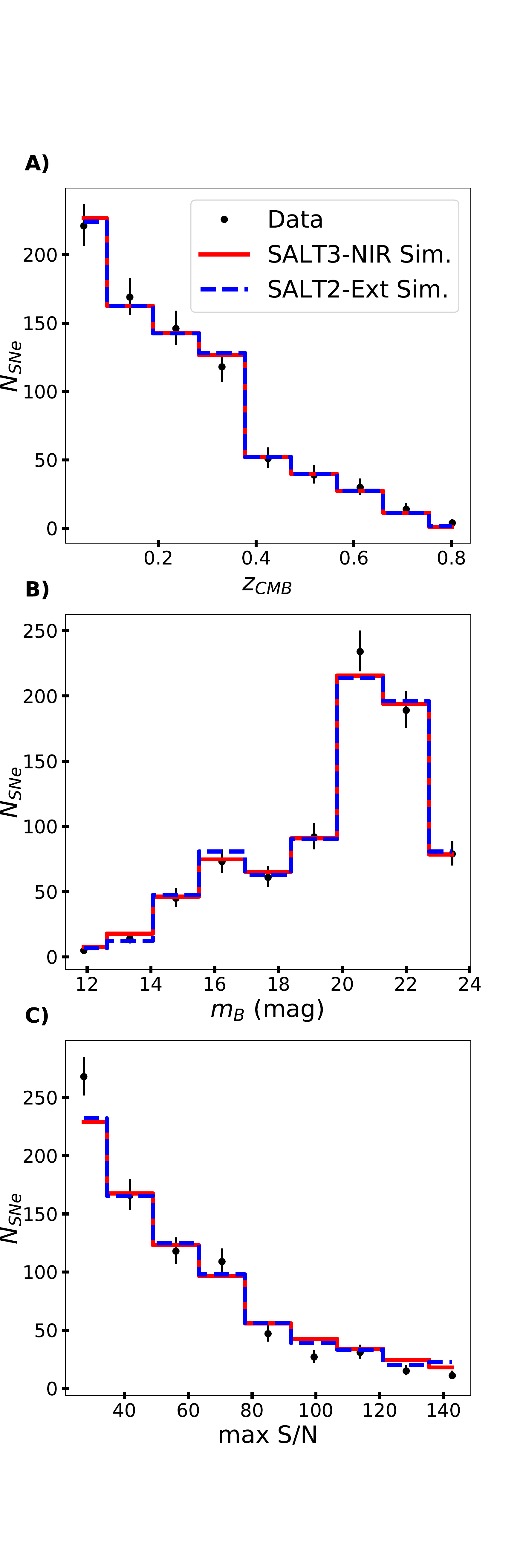}
    \caption{Distributions of redshift (top), SALT2-$m_B$ (middle), and maximum S/N (bottom) for training sample data (filled circles), SNANA simulation with SALT3-NIR model (red histogram), and simulation with SALT2-Extended model (blue histogram).}
    
    \label{fig:sim_data}
\end{figure}

\subsection{Producing Simulated Training Samples}
\label{sub:val_sims}
To validate the training procedure, we simulate samples of light curves and spectra with observational characteristics [i.e., cadence, signal-to-noise ratio (S/N), and filter/wavelength coverage] matching the training sample described in Section \ref{fig:p22_sample}. We employ the widely used simulation in the SuperNova ANAlyis (\texttt{SNANA}) software package \citep{kessler_snana:_2009}, which generates realizations of SN photometry following Figure 1 of \citet{kessler_first_2019}. Briefly, the simulation begins with a rest-frame ``Source SED" and propagates the SN light through an expanding universe, the Milky Way, Earth’s atmosphere, and instrumental filters, finally generating CCD photoelectrons. To produce realistic spectra, the simulation creates Monte Carlo realizations of a source SED using an empirical model of S/N vs. wavelength derived from the training sample described above. These simulations were shown to accurately reproduce both photometry and spectra for the \citetalias{kenworthy_salt3_2021} simulated training sample. 

The choice of model for simulating light curves and spectra is in principle arbitrary, as \saltshaker is capable of training a model regardless of the underlying simulation framework (Dai et al., in prep.). However, in order to directly use $M_0$ and $M_1$ recovery precision as a training accuracy metric, it is necessary to select a model described by Equation \ref{eq:salt}.  We therefore make two choices to evaluate the accuracy of our model. First, we use the model produced in Section \ref{sub:training} as the source SED, which implicitly assumes that the trained SALT3-NIR model is a perfect description of SNe\,Ia; this serves as a floor for the validation stage. Second, we produce an identical set of simulations using SALT2-Extended (Section \ref{sub:salt2_ext}), which assumes the same formalism as SALT3-NIR but has poorly trained color and stretch relationships as well as a model surface that varies on much smaller wavelength scales compared to SALT3-NIR \citep[see Figure \ref{fig:s3nir_salt2ext} and][]{pierel_extending_2018}. As \saltshaker is designed to ensure a smoothly varying model in phase and wavelength, we expect this scenario to result in a worse training performance. However, this training is a useful exercise as it gives a ceiling for training accuracy, in the case where the frequency of true variability is higher than the level \saltshaker is designed to encode into the final model surface. 

As \citetalias{kenworthy_salt3_2021} have already confirmed that \saltshaker accurately produces a model given a simulated training sample with realistic \SNIa population parameters, here we check that \saltshaker recovers the underlying NIR model components when training over the updated wavelength range. We therefore produce simulated light curves with the same cadence, filter coverage, number of observations, redshift, and S/N as the true training sample, as well as the same number of spectra with matching wavelength coverage. We do not match the model parameter distributions of $x_1$ and $c$ identically to those of the training sample (as \citetalias{kenworthy_salt3_2021} did), but instead sample Gaussian distributions that encompass the full range in expected values for these parameters (i.e., $\mu_{x_1}=0,\sigma_{x_1}=1$ and $\mu_c=0,\sigma_c=0.1$). The resulting simulation enables us to check if \saltshaker is robust in recovering model surfaces over the entire plausible range in model parameters, given the same volume of data as the training sample in Section \ref{sub:full_sample}. 

\begin{figure}
    \centering
    \includegraphics[trim={1.5cm 0cm 1.9cm 1cm},clip,width=\linewidth]{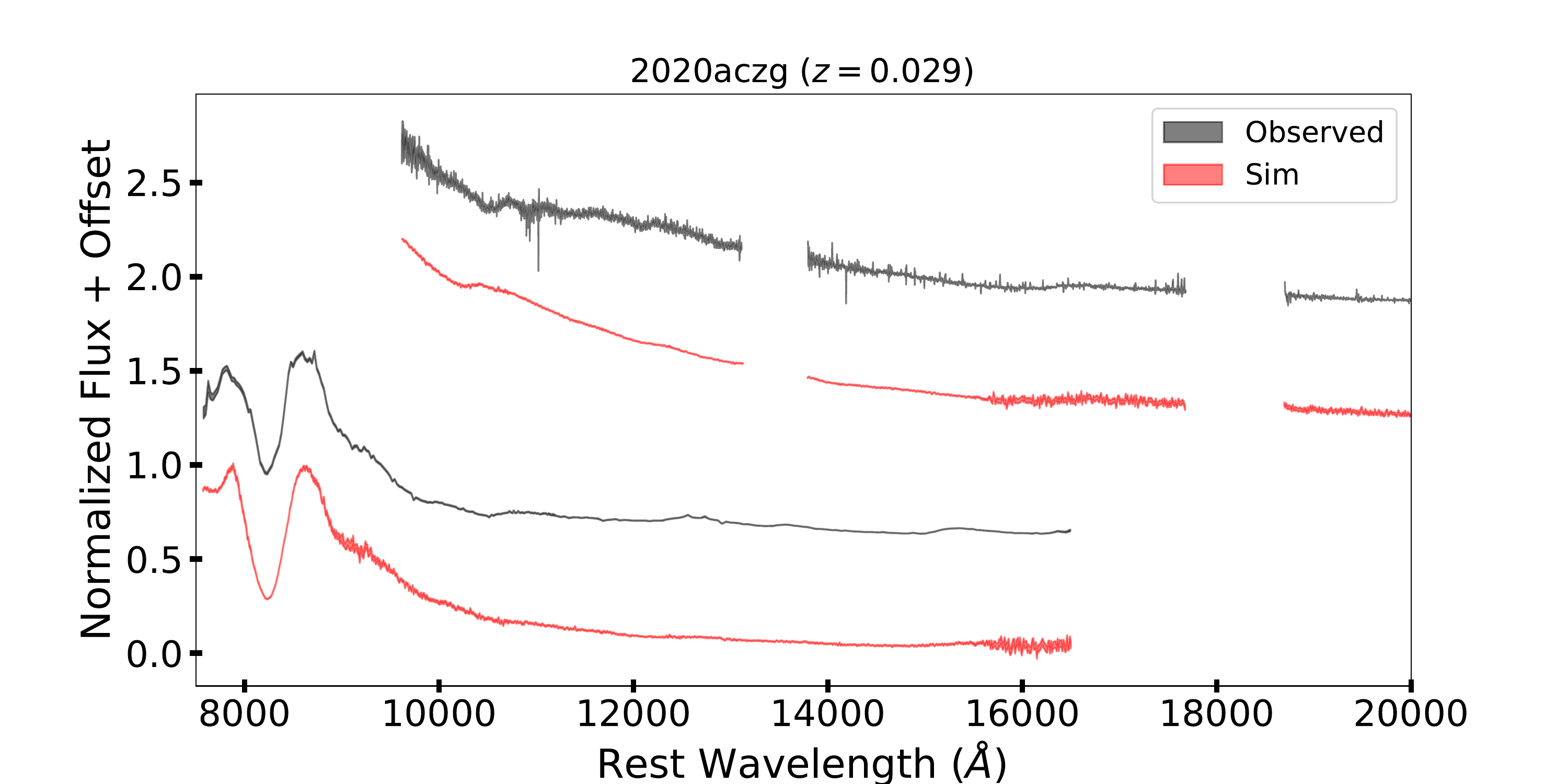}
    \caption{Example of observed (black) and simulated (red) NIR spectra for SIRAH SN\,2020aczg. The lower spectra are from the WFC3/IR grism, while the upper are from Keck NIRES. The gaps in the NIRES spectra correspond to atmospheric absorption windows. Every spectrum has a random offset applied to visually distinguish them.}
    \label{fig:nir_spec}
\end{figure}

We create 10 distinct realizations of each simulated training set to ensure sufficient statistical sampling, and  Figure \ref{fig:sim_data} displays the agreement of key parameters between randomly selected simulations and the training sample from Section \ref{sec:data}. While the simulations are not expected to be a perfect representation of the data, these key distributions of redshift, $m_B$, and maximum S/N are all sufficiently well-matched to accept the simulations as appropriate representations of the training sample. An example of observed and simulated NIR spectra is shown in Figure \ref{fig:nir_spec}.

\subsection{Training on Simulations with \texttt{SALTShaker}}
\label{sub:train_sim}
 
\begin{figure}[h!]
    \centering
  
    \includegraphics[trim={0cm 0cm 0cm 0cm},clip,width=\linewidth]{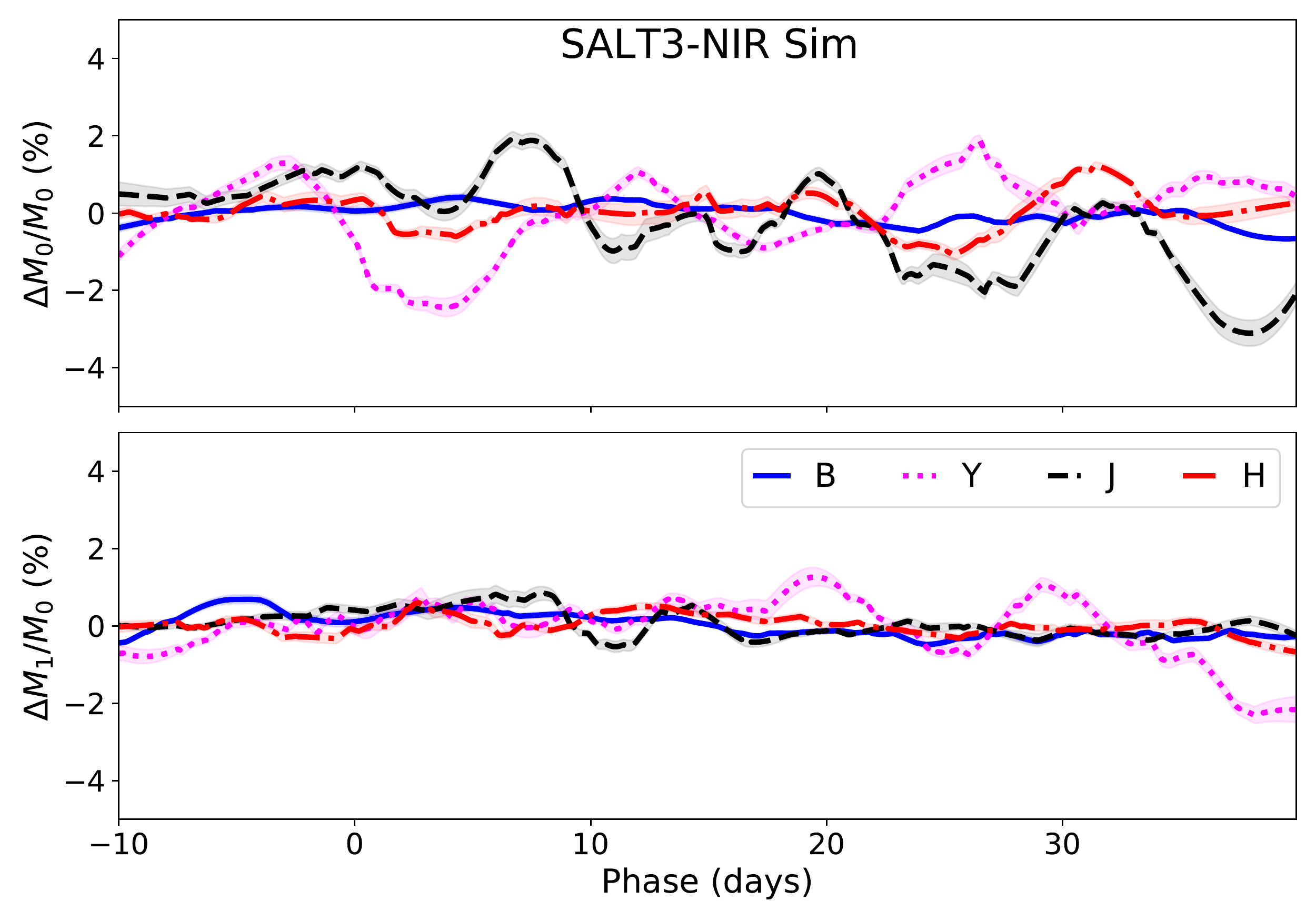}
    \caption{Result of the SALT3-NIR simulation. Difference between the simulated and trained $M_0$ (top) and $M_1$ (bottom) model components, integrated over the $BYJH$ bandpasses, and divided by the $M_0$ component to show fractional differences. Each curve is the average of 10 training sets simulated using SALT3-NIR as the source SED. The $B$ band is shown as a target model fidelity for cosmology. }
    \label{fig:phase_m0_m1_comp}
\end{figure}

We train a version of SALT3-NIR on each of the simulated training samples
described in Section \ref{sub:val_sims}. After integrating the simulated and trained $M_0$ and $M_1$ components over the $YJH$ bandpasses, we compare the recovered model surfaces to those input in the simulation. The SALT3-NIR simulation is an optimistic scenario, in which the training process is perfectly suited to the data. As a result, we find that the $M_0$ component is constrained to within $\sim2\%$ across nearly the full phase range. \saltshaker also recovers the $M_1$ component to within $\sim1\%$ over the same phase range, despite present phase gaps in the NIR training sample (Figure \ref{fig:phase_m0_m1_comp}). This is due in part to the relatively small $M_1$ amplitude compared to the $M_0$ component (Figure \ref{fig:s3nir_salt2ext}).

\begin{figure}[h!]
    \centering
    \includegraphics[width=\linewidth]{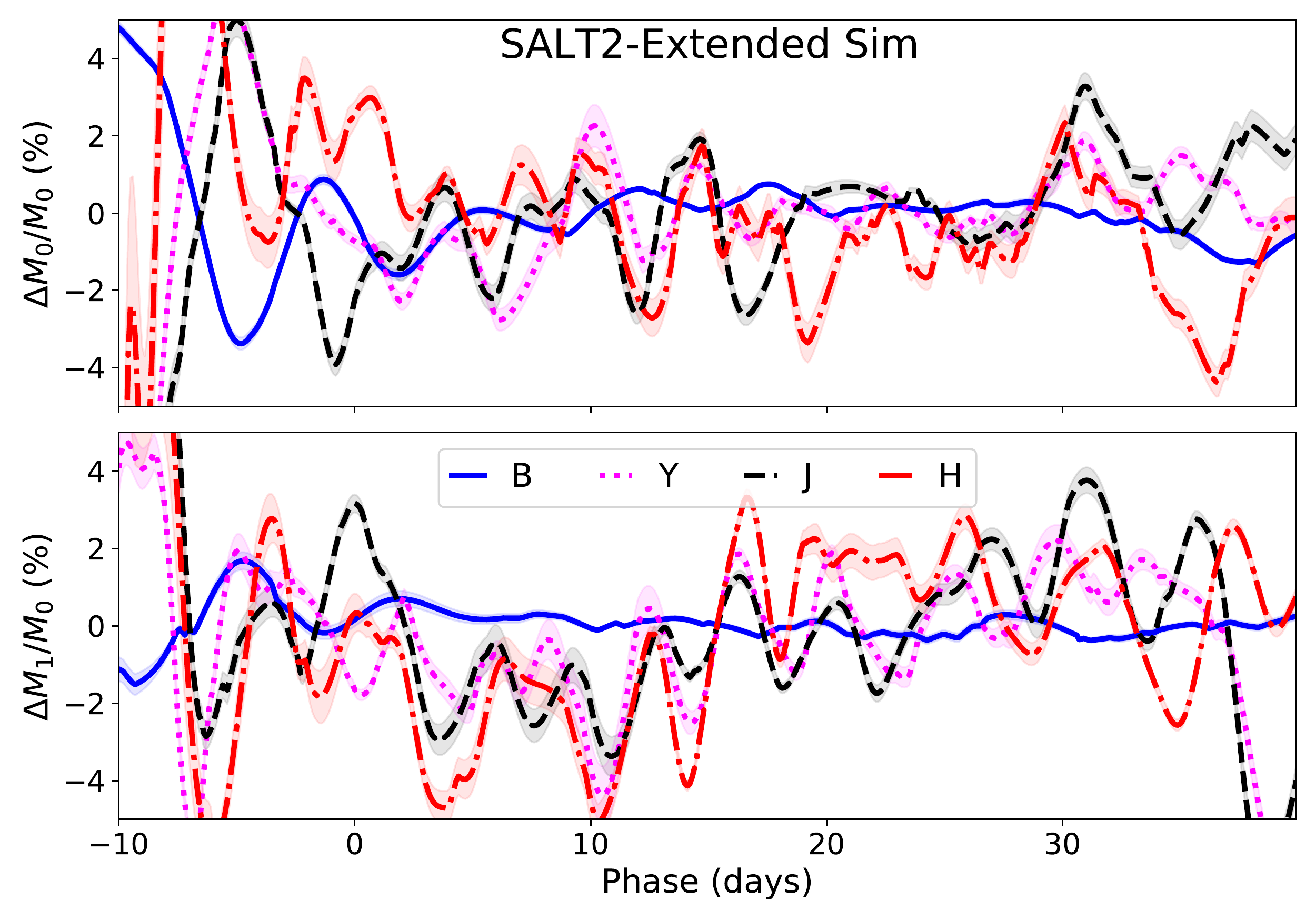}
    
    \caption{Result of the SALT2-Extended simulation. Difference between the simulated and trained $M_0$ (top) and $M_1$ (bottom) model components, integrated over the $BYJH$ bandpasses, and divided by the $M_0$ component to show fractional differences. Each curve is the average of 10 training sets simulated using SALT2-Extended as the source SED. The $B$ band is shown as a target model fidelity for cosmology.}
    \label{fig:salt2_phase_m0_m1_comp}
\end{figure}

On the other hand, the SALT2-Extended simulation is the ``pessimistic'' scenario, in the sense that its behavior has spurious variability beyond what \saltshaker is attempting to represent. Nevertheless, Figure \ref{fig:salt2_phase_m0_m1_comp} shows that \saltshaker is still capable of retrieving the $M_0$ model component to within $\sim2$--3\% over much of the phase range. The $M_1$ component is more poorly retrieved, likely because the high variability in SALT2-Extended is treated as signal by \saltshaker, which is allocated to $M_1$. Shown for comparison in Figures \ref{fig:phase_m0_m1_comp} and \ref{fig:salt2_phase_m0_m1_comp} is the recovery of the $B$-band model flux for the same simulation and training processes. Both model components are recovered to within 1\% across the full phase range in the optical (apart from $\lesssim -5$\,days), setting a fidelity target for the NIR to be reliably used for cosmology. 

\subsection{Comparison with SALT3}
\label{sub:salt3_val}
The differences in optical color scatter shown in Figure \ref{fig:salt3-nir_errs}b demonstrate that simultaneously training the optical and NIR slightly degrades the optical model performance. Some variation in scatter is expected, as the model surfaces are continuous in wavelength, but the cause of this degradation is not understood. The most likely culprits are either (a) the NIR is not nearly as well-constrained as the optical owing to a lack of sufficient data, and therefore the NIR scatter impacts optical wavelengths, or (b) the NIR flux is more uniform compared with the optical but \saltshaker is only able to weight individual filters instead of the optical as a whole, which increases the color scatter at optical wavelengths. While the following section suggests that the latter hypothesis is correct, the SALT3 optical-only model should be used in \SNIa cosmology analyses until this optical+NIR scatter issue is understood.

Nevertheless, here we check that optical distances measured with SALT3-NIR are comparable to those of SALT3 by fitting the 10 SALT2-Extended simulated training samples from Section \ref{sub:train_sim} with both models, using optical data only and the methods described in Section \ref{sub:salt_dist}. The results are summarized in Tables \ref{tab:sim_ab_rms} and \ref{tab:salt3_mu}. The $\alpha$ ($\beta$) parameter shown is the coefficient of relation between SN\,Ia luminosity and stretch (color), shown in Equation \ref{eq:tripp}. The two models agree within 3$\sigma$ (or $<2$\%) for $\beta$, but a significant difference in $\alpha$ ($\sim20$\%) suggests that the NIR training process impacts the overall stretch-luminosity correlation. Since we ignore distance-bias corrections (Section \ref{sub:salt_dist}), the absolute values of $\alpha$ and $\beta$ should not be compared with other studies, but are instead intended for direct comparison between SALT3 and SALT3-NIR. Although there are differences in the measured $\alpha$ and $\beta$ parameters, we find that the slope of the $x_1$ ($c$) correlation from measurements with each model is 1.04 (0.99) with a root mean square error (RMSE) of 0.13 (0.02), meaning light-curve parameter measurements are essentially identical for the two models. The Hubble residual root-mean square (RMS) is also not significantly different between the two models. 

Table \ref{tab:salt3_mu} shows that differences in optical distance measurements are $\lesssim0.01$\,mag, meaning that SALT3-NIR is reliable for optical+NIR light-curve fitting or simulations. Here $\Delta\mu$ is the binned residual between measured distances for every simulated \SNIa, and in both tables the uncertainties are the standard error on the mean (SEM) calculated from the 10 simulated training sets.

\begin{table}[h]
    \caption{\label{tab:sim_ab_rms} $\alpha,\ \beta$, and Hubble residual RMS measured using SALT3 and SALT3-NIR, and the 10 training sets simulated with SALT2-Extended.}
    \begin{tabular*}{\linewidth}{@{\extracolsep{\stretch{1}}}*{3}{c}}\toprule
    Parameter&SALT3&SALT3-NIR\\
    \hline
    $\alpha$&0.133$\pm$0.003&0.105$\pm$0.003\\
    $\beta$& 2.846$\pm$0.017&2.888$\pm$0.017\\
    HR$_{\rm{RMS}}$ (mag)&0.117$\pm$0.002&0.118$\pm$0.002\\
    \end{tabular*}

\end{table}

\begin{table}[h]
\caption{\label{tab:salt3_mu} Binned differences in optical distance measurements between SALT3 and SALT3-NIR.}
\begin{tabular*}{\linewidth}{@{\extracolsep{\stretch{1}}}*{3}{c}}
\toprule
 $z$ bin&$\Delta\mu$&N$_{\rm{SN}}$ \\
 &(mag)&\\
\hline
0.015--0.026&$0.004\pm0.001$&78\\
0.026--0.047&$0.002\pm0.001$&78\\
0.047--0.082&$0.000\pm0.001$&61\\
0.082--0.146&$0.001\pm0.001$&80\\
0.146--0.257&$0.000\pm0.001$&235\\
0.257--0.453&$0.005\pm0.001$&265\\
0.453--0.800&$0.000\pm0.001$&125\\
\end{tabular*}
\end{table}

\subsection{Hubble Residuals}
\label{sub:hr}
Finally, we examine the impact of the NIR in accurately standardizing SNe\,Ia. First, we take the subset of 105 \SNeIa in our training sample (Section \ref{sec:data}) that have sufficient rest-frame $YJH$ data for light-curve fitting without optical bands (i.e., at least 3 data points in at least 1 NIR filter), and are at a sufficiently high redshift such that the effects of peculiar velocities are mitigated (i.e., $z>0.015$). We fit the optical ($BVRI$) light curves in this sample with SALT3, and subsequently fix these time of peak estimates for NIR-only fitting. We reject \SNeIa with optically-fitted parameters $|c|>0.3$ or $|x_1|>3$ \citep[e.g.,][]{scolnic_complete_2018}. SALT3 and SALT3-NIR distance measurements are derived following Section \ref{sub:salt_dist}. In \textbf{all} cases below the results shown for SALT3 are using only optical fitting and the full set of SALT3 parameters, resulting in RMS values of 0.16--0.19\,mag. 

In addition to optical SALT3 constraints, we compare SALT3-NIR distance measurements to distances determined using the ``SNooPy'' \texttt{ebv\_model(2)} \citep{burns_carnegie_2011,burns_carnegie_2014,avelino_type_2019,jones_cosmological_2022}. SNooPy has been effectively the only SN\,Ia light-curve model capable of reliably fitting the NIR for much of the last decade\footnote{We use the SNANA implementation of SNooPy \texttt{ebv\_model(2)} light-curve fitting.}, and was trained on well-calibrated CSP light curves \citep{burns_carnegie_2011,krisciunas_carnegie_2017}. The more recent BayeSN model \citep{mandel_hierarchical_2020} also extends to the NIR and uses a different light-curve fitting algorithm. However, the fitting code is not currently open source, and we therefore restrict our comparison to SNooPy.

We split the NIR sample using all combinations of rest-frame NIR filters ($Y$, $J$, $H$, $YJ$, $YH$, $JH$, $YJH$), and fit each subsample with both SALT3-NIR and SNooPy. For each case, we fit with only the light-curve amplitude (i.e., $x_0$ for SALT3-NIR and luminosity distance for SNooPy), and separately fit the amplitude with model stretch ($x_1,s_{BV}$). For subsamples with 2 or more NIR filters, we attempt to include a color parameter ($c, E(B-V)$). We fit with relaxed bounds of $|c|<1, |x_1|<4$ for SALT3-NIR and $|E(B-V)|<1.3$\,mag, $0.7<s_{BV}<1.3$ for SNooPy \citep[with $R_V=1.518$;][]{burns_carnegie_2014}. 

\begin{figure*}[h!]
    \centering
    \includegraphics[width=\textwidth]{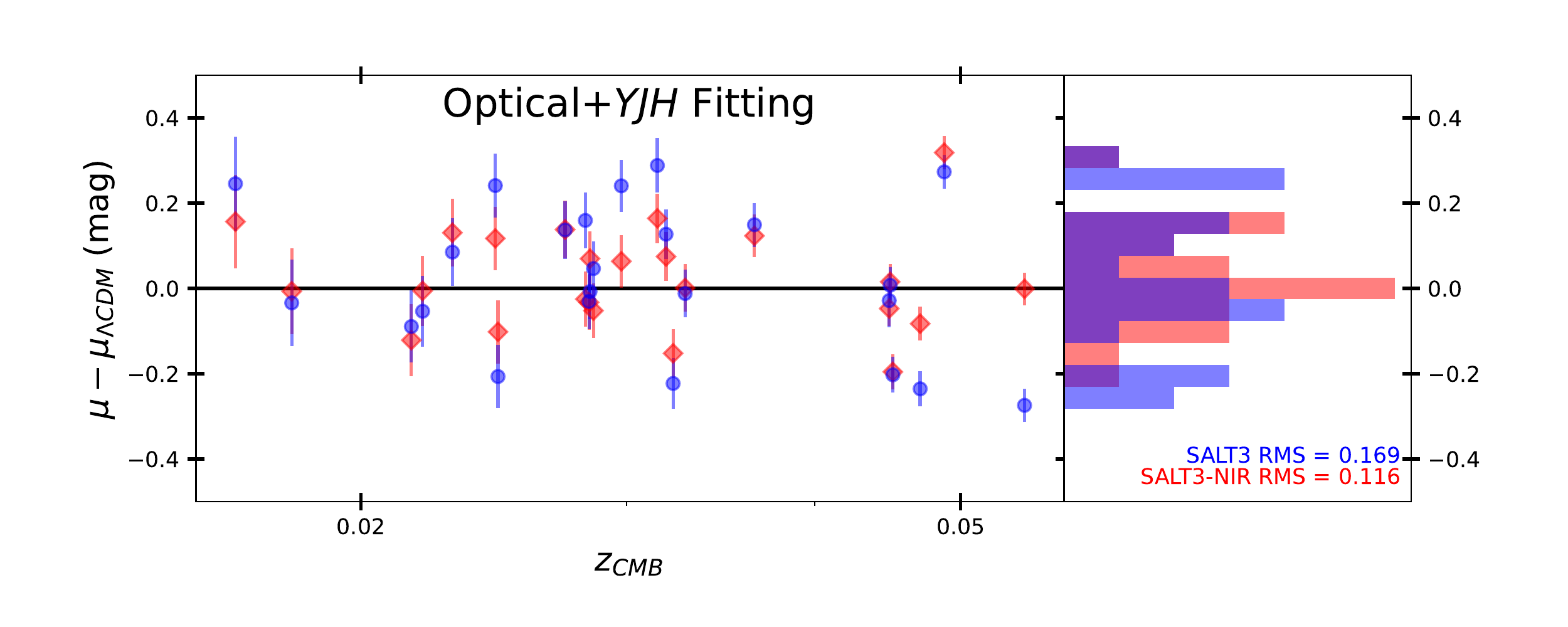}
    \caption{Hubble residuals (with respect to $\Lambda$CDM) when fitting optical+NIR filters with SALT3-NIR (red diamonds), and optical filters with SALT3 from \citetalias{kenworthy_salt3_2021} (blue circles). Here, all model parameters are allowed to float.}
    \label{fig:opt_nir_hr}
\end{figure*}

\begin{figure*}[h!]
    \centering
    \includegraphics[width=\textwidth]{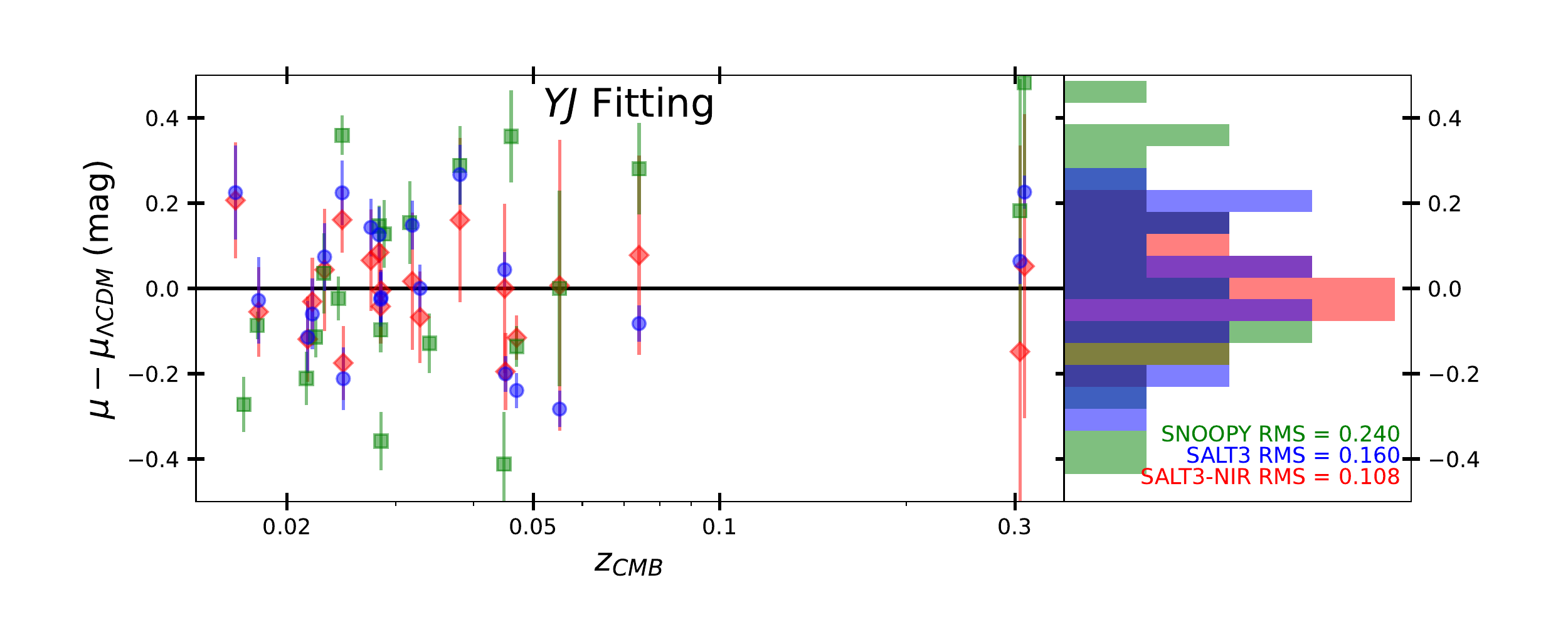}
    \caption{Hubble residuals (with respect to $\Lambda$CDM) when fitting only rest-frame $YJ$ filters with SALT3-NIR (red diamonds) and SNooPy (green squares), and optical filters with SALT3 from \citetalias{kenworthy_salt3_2021} (blue circles). Here, the stretch parameter is not allowed to float for SALT3-NIR and SNooPy.}
    \label{fig:nir_only_hr}
\end{figure*}

Table \ref{tab:rms_comp} shows the results of this fitting process. The SALT3 RMS values correspond to the optical RMS for the respective subsample, while the SALT3-NIR and SNooPy RMS values are the result of fits to the subsample using the given set of parameters. Both SALT3-NIR and SNooPy use flat priors for all parameters, and in cases where stretch (color) is not fit we fix the value to $x_1=0$ ($c=0$) for SALT3-NIR and $s_{BV}=0.9$ ($E(B-V)=0$\,mag) for SNooPy. This choice is based on \citet{jones_cosmological_2022}, who also found that shifting the fixed parameter value had a negligible impact on Hubble residual RMS. Figure \ref{fig:opt_nir_hr} shows the optical+NIR case, while the scenario with the best SALT3-NIR RMS is shown in Figure \ref{fig:nir_only_hr}. We cannot fit the optical+$YJH$ case with SNooPy, as the optical data for a large number of the \SNeIa in this sample come from ATLAS, whose wide bandpasses cause known issues with the K-corrections necessary for SNooPy fitting.

The numbers in each Table \ref{tab:rms_comp} section correspond to results for the \SNeIa passing cuts for all subsamples in that section. For example, in the first section there are 44 \SNeIa that pass all of the $Y$, $J$, and $H$ amplitude-only fitting procedures, and the RMS values shown (i.e., $\rm{RMS}_Y=0.136, \rm{RMS}_J=0.161, \rm{RMS}_H=0.145$\,mag for SALT3-NIR) are based on those identical 44 \SNeIa. This procedure enables a direct comparison between results in a given section, but significantly reduces the sample size in each row. For each amplitude-only fitting procedure, where far fewer fits fail, we show the RMS for all \SNeIa passing that single fit in parentheses. For example while only 26 \SNeIa pass all 4 fitting procedures in the $YJ$ section, 54 \SNeIa are successfully fit when floating only the amplitude. In this scenario, SNooPy produced the lowest RMS while fitting the amplitude only for the sample of 26 \SNeIa, but SALT3-NIR performs
the best on the full sample of 54 \SNeIa with sufficient $YJ$ data.

\begin{table*}
\caption{\label{tab:rms_comp} Hubble residual RMS comparison between SALT3-NIR, SNooPy, and SALT3.}
\begin{tabular*}{\linewidth}{@{\extracolsep{\fill}}llrrrr}
\toprule
 &&\multicolumn{3}{c}{Hubble Residual RMS for:}&\\
 Filters&Parameters&SALT3-NIR&SNooPy&SALT3$^a$ & $N_{\rm{SN}}^b$ \\
 &&mag (mag)&mag (mag)&mag (mag)&\\
\hline
$Y$&Amplitude$^c$&0.164 (0.151)$^d$&0.190 (0.183)&0.174 (0.135)&31 (74)\\
$J$&Amplitude&0.169 (0.178)&0.187 (0.207)&0.174 (0.134)&31 (47)\\
$H$&Amplitude&0.161 (0.158)&0.159 (0.156)&0.174 (0.163)&31 (38)\\
\hline
$Y$&Amplitude, Stretch$^e$&0.142&0.172&0.158&28\\
$J$&Amplitude, Stretch&0.170&0.187&0.158&28\\
$H$&Amplitude, Stretch&0.135&0.122&0.158&28\\
\hline
$YJ$&Amplitude&\textbf{0.117}$^g$ (0.154)&0.111 (0.199)&0.160 (0.136)&21 (46)\\
$YJ$&Amplitude, Color$^f$&\textbf{0.108}&0.240&0.160&21\\
$YJ$&Amplitude, Stretch&\textbf{0.115}&0.116&0.160&21\\
$YJ$&Amplitude, Stretch, Color&\textbf{0.118}&0.274&0.160&21\\
\hline
$YH$&Amplitude&0.151 (0.160)&0.143 (0.160)&0.160 (0.152) &32 (39)\\
$YH$&Amplitude, Color&0.149&0.175&0.160&32\\
$YH$&Amplitude, Stretch&0.151&0.151&0.160&32\\
$YH$&Amplitude, Stretch, Color&0.153&0.194&0.160&32\\
\hline
$JH$&Amplitude&0.138 (0.159)&0.141 (0.155)&0.117 (0.160)&21 (38)\\
$JH$&Amplitude, Color&0.116&0.151&0.117&21\\
$JH$&Amplitude, Stretch&0.158&0.109&0.117&21\\
$JH$&Amplitude, Stretch, Color&0.129&0.160&0.117&21\\
\hline
$YJH$&Amplitude&\textbf{0.128} (0.157)&0.116 (0.183)&0.169 (0.162)&24 (37)\\
$YJH$&Amplitude, Color&\textbf{0.126}&0.184&0.169&24\\
$YJH$&Amplitude, Stretch&\textbf{0.126}&0.108&0.169&24\\
$YJH$&Amplitude, Stretch, Color&\textbf{0.131}&0.181&0.169&24\\
$OPT+YJH$&Amplitude, Stretch, Color&\textbf{0.116}&--&0.169&24\\
\hline

\end{tabular*}

$^a$Calculated using optical fits (with all SALT3 parameters) to the respective subsample.

$^b$Refers to the number of \SNeIa that have the rest-frame filters listed and pass fitting cuts.

$^c$Corresponds to the $x_0$ parameter for SALT3-NIR and luminosity distance for SNooPy.

$^d$Numbers in parentheses correspond to the results for all SNe passing the amplitude-only fitting in that row.

$^e$Corresponds to the $x_1$ parameter for SALT3-NIR and $s_{BV}$ for SNooPy.

$^f$Corresponds to the $c$ parameter for SALT3-NIR and $E(B-V)$ for SNooPy.

$^g$Bold numbers correspond to cases where the SALT3-NIR RMS is lower than SALT3 with $> 95$\% confidence
\end{table*}

\begin{figure*}
    \centering
    \includegraphics[width=\linewidth,trim={.5cm .5cm 0cm 0cm},clip]{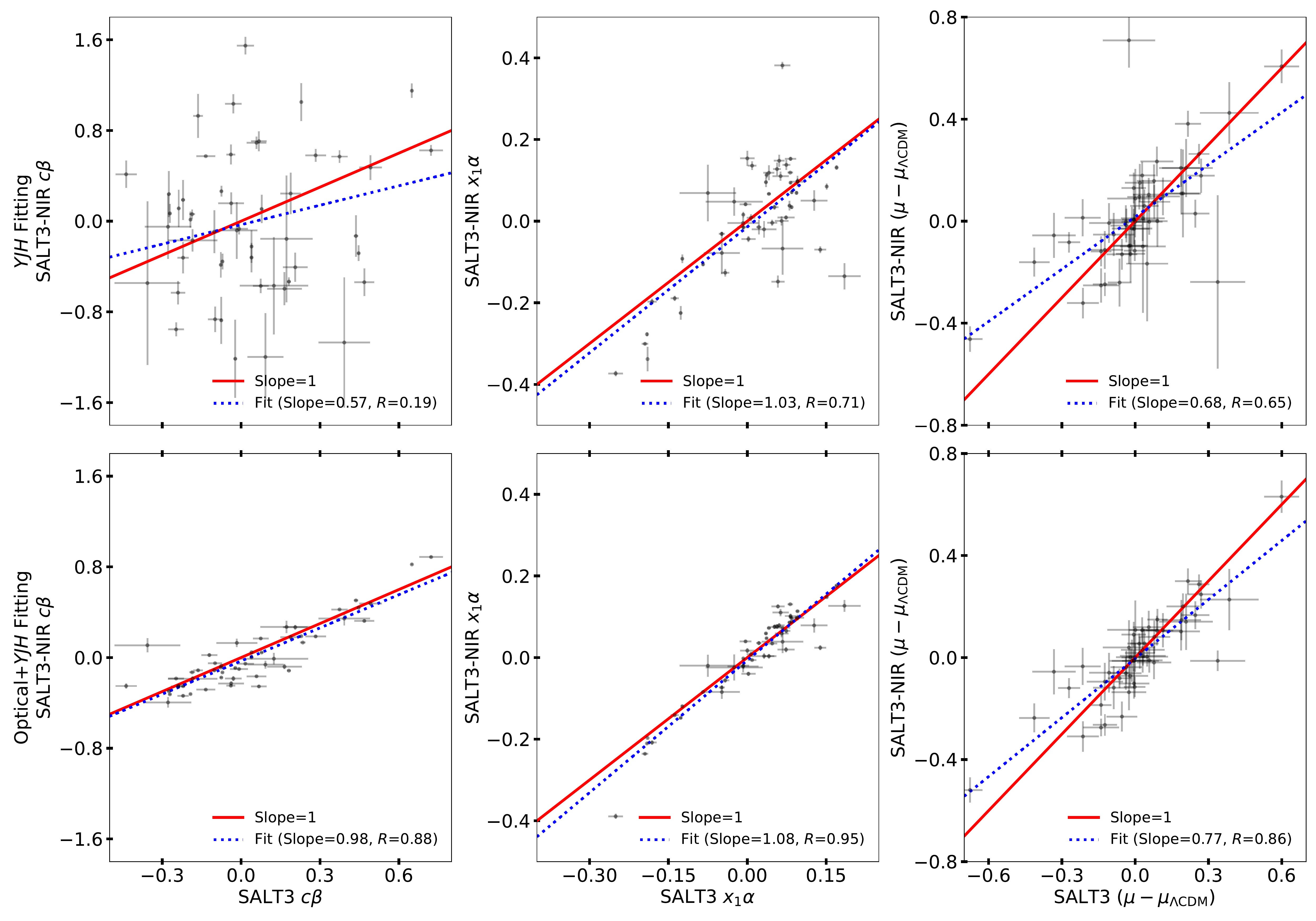}
    \caption{\label{fig:fitting_corrs} Correlations (black points with error bars) between SALT3 and SALT3-NIR measured values of $c$ (left), $x_1$ (center), and $\mu-\mu_{\Lambda\rm{CDM}}$ (right) for $YJH$ (top row) and Optical$+YJH$ (bottom row) fitting scenarios. The SALT3 fits are for the same \SNeIa fit by SALT3-NIR, but using only optical data. Best-fit linear regressions are shown as blue dotted lines, and a slope of 1 (red solid line) is given for comparison.}
    
\end{figure*}


The SALT3-NIR RMS is lower than SNooPy in 20 of 29 scenarios, and only one scenario is more than 10\% higher. Optical distances with SALT3 only produce the lowest RMS in 4 of the 29 scenarios, with the largest difference being the $JH$-band fitting (10--$30$\%). In the best SALT3-NIR case ($YJ$, fitting the amplitude and color parameters), we see a $\sim30$\% reduction in residual RMS compared to optical fitting, with a similar effect seen when fitting optical and NIR data simultaneously.  Although the sample here is small, we attempt to characterize the significance of this result by bootstrapping the NIR-only sample. We define a new set of parameters $\Delta\mu_i$, such that
\begin{equation}
    \label{eq:boot}
    \Delta\mu_i = \delta\mu_{i,\rm{NIR}}^2-\delta\mu_{i,\rm{Optical}}^2,
\end{equation}
where $\delta\mu_i$ refers to the Hubble residual for the $i^{th}$ \SNIa. We bootstrap the set of $\Delta\mu_i$ values for every row in Table \ref{tab:rms_comp} with 10,000 samples, and find that fitting the NIR data alone provides a lower RMS than optical fitting in 8 scenarios with 95\% confidence. We also find that optical+NIR fitting produces a lower RMS than optical-only fitting with 95\% confidence. This is in agreement with \citet{avelino_type_2019}, who found that using the NIR, alone or with optical data, resulted in a significant improvement in Hubble residual RMS. However, this is the first attempt at applying stretch and/or color corrections at NIR wavelengths, with these initial results suggesting that stretch corrections at least are certainly worthwhile in the current framework. All scenarios in which the bootstrapping analysis confirmed the SALT3-NIR RMS is lower than SALT3 with $>95$\% confidence are bold in Table \ref{tab:rms_comp}.

Figure \ref{fig:fitting_corrs} shows correlations between SALT3 and SALT3-NIR for measured values of $c,x_1$, and $\mu-\mu_{\Lambda\rm{CDM}}$ for the $YJH$ and Optical$+YJH$ fitting scenarios described above. Values measured by SALT3 are still using optical data only, for the same \SNeIa fit by SALT3-NIR. As expected, when optical data are included in SALT3-NIR fitting, the correlation between SALT3 and SALT3-NIR for each parameter is quite strong, with a slope close to 1 for both $c$ (0.98) and $x_1$ (1.08). Interestingly, there is still a strong correlation between measurements of $x_1$ when optical data are excluded from the SALT3-NIR fit (slope = 1.03, $R=0.71$), though this relationship should be investigated further as it appears driven by a relatively small number of points. Nevertheless, this suggests that the relationship between light-curve stretch and luminosity impacts the NIR as well as the optical, and that the NIR is capable of measuring this relationship on its own. This is supported by consistent improvements in Hubble residual RMS when a stretch parameter is included in NIR-only fitting, shown in Table \ref{tab:rms_comp}. On the other hand, attempts to measure the $c$ parameter using the NIR alone display very little correlation with optical measurements. This is likely due to the inherent difficulty in measuring color variation in the NIR (e.g., a difference of $\Delta c$ = 0.1 changes the $J-H$ color by 0.003\,mag and the $B-V$ color by 0.1\,mag), but future work should investigate the possibility that this result could be hinting at a different physical origin for NIR color variation compared to optical. We find that in cases where the difference between optical and NIR measurements of $c$ are large, there is a correspondingly large and opposite difference in measurements of $x_1$ that lead to a relatively small change in distance. 

\section{Conclusion}
\label{sec:conclusion}
We have used the open-source Python package \saltshaker to extend the training of SALT3 to $\maxmodelwave\,\mu$m. The training sample has been left unchanged in the optical, apart from optical counterparts to the new NIR sample, so that SALT3-NIR is reliable across its full wavelength range, with only slightly higher color scatter in the optical. Future efforts to improve the calibration in the additional training sample, at both optical and NIR wavelengths, would increase the fidelity of the model. Using simulated training samples, we have shown that SALT3 and SALT3-NIR optical distances differ by $<0.01$\,mag, and that \saltshaker is able to constrain the source SED to within $\sim 2$--3\% in the NIR given our data density, though $\lesssim1\%$ is the target for a robust cosmological light-curve fitter. 

Still, with a sample of low-$z$ SNe having rest-frame NIR photometry, we find that using SALT3-NIR produces comparable single-filter distance measurements relative to the SNooPy light-curve fitter (but improved in $Y,J$). When adding multiple filters or fitting stretch and/or color parameters, SALT3-NIR provided NIR-only Hubble residual RMS values $\sim0.02$-0.03\,mag lower than SNooPy, reaching $<0.11$\,mag, which is a $\sim30\%$ improvement over SALT3 using optical filters for the same \SNeIa. We also find that Hubble residual scatter is lower when using SALT3-NIR and only NIR data than optical distances in 90\% of our fitting scenarios (8 scenarios at $>95$\% confidence), and that the RMS from optical+NIR light-curve fitting is also lower than optical alone with 95\% confidence. These results are in agreement with those of \citet{avelino_type_2019}, who used an amplitude-only fitting procedure, but we find that stretch and color corrections provide a $\sim10\%$ improvement when fitting $JH$ filters and can result in a similar RMS compared with optical+NIR fitting. Overall, we find that NIR and optical stretch measurements correlate well while color measurements are uncorrelated, though both luminosity corrections often improve NIR distance measurements (by up to $\sim20\%$ for $JH$ filters with a color correction). More work must be done to understand if the difference between optical and NIR color variation has a physical origin, or simply requires innovations in the modeling or training process to fully harness.

The indication that the NIR alone can provide a lower RMS than optical+NIR, and the large difference in $\alpha$ seen when fitting optical light curves with both SALT3 and SALT3-NIR, suggest that future work should investigate whether adding a new model component to SALT specifically for the NIR would improve distance measurements. The lack of correlation between color measured with optical and NIR filters is likely another indicator of this need. \citet{mandel_hierarchical_2022} found that optical and NIR bands were relatively uncorrelated, which could be providing similar evidence for this result. Simply allowing uncertainties in color scatter to be correlated could be sufficient, and so this concept should also be explored. Significantly more data would be necessary for this investigation, but it should be possible using \roman. 

\begin{figure}[h!]
    \centering
    \includegraphics[trim={0cm 0cm 1.5cm 1.5cm},clip,width=\linewidth]{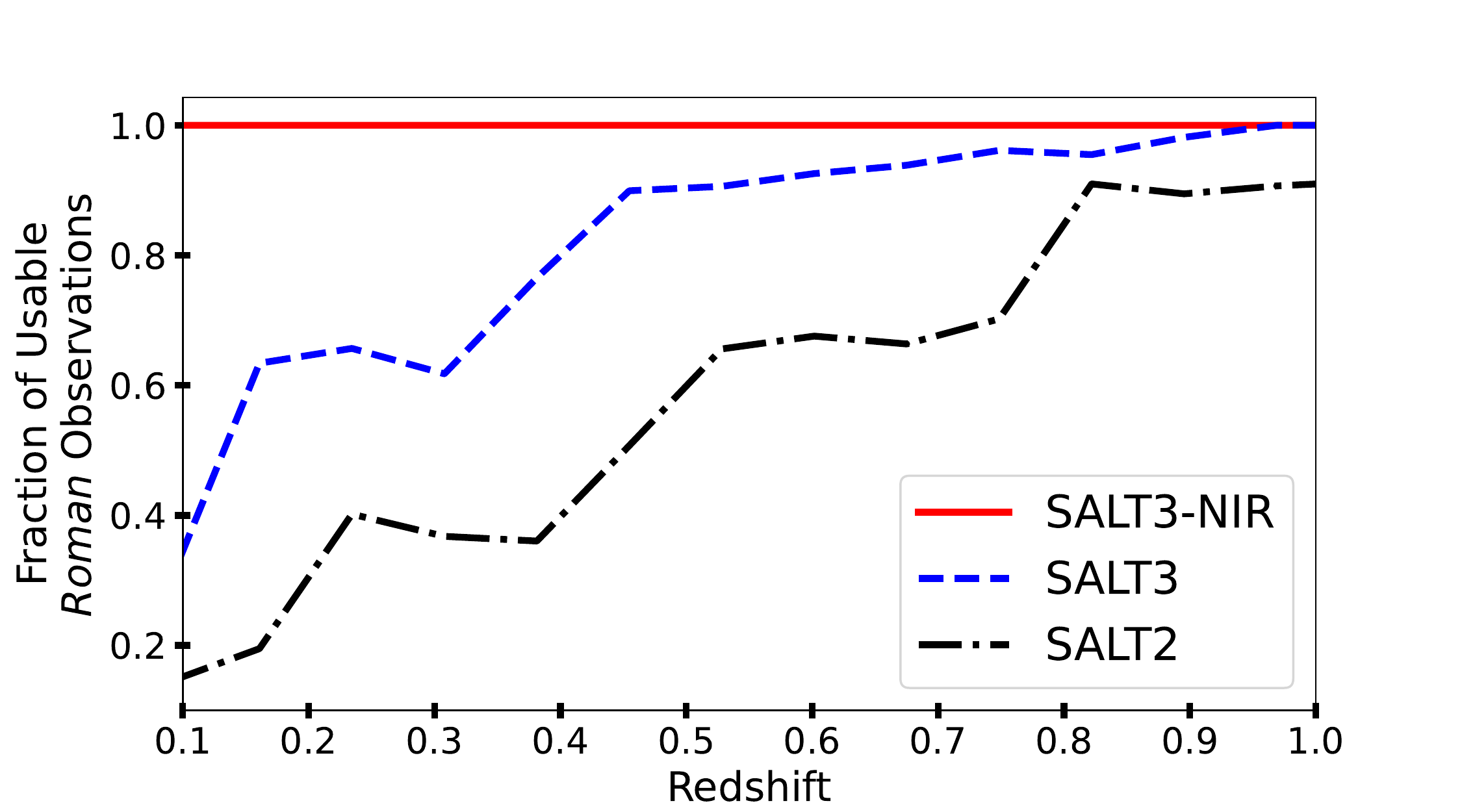}
    \caption{The fraction of all \textit{Roman} HLTDS observations that SALT2 (black dash-dot), SALT3 (blue dashed), and SALT3-NIR (red solid) are capable of fitting (i.e., the wavelength range fully encompasses the respective filter) as a function of redshift. The difference is most significant at low redshifts, where \textit{Roman} filters probe the rest-frame NIR. }
    \label{fig:roman_obs}
\end{figure}

The new wavelength range of SALT3-NIR will enable the analysis of significantly more \SNIa data with \roman, all at $z\lesssim1$ in the rest-frame NIR where SNe\,Ia have been shown repeatedly to be better standard candles \citep[e.g.,][and see Section \ref{sub:hr}]{mandel_type_2011,dhawan_measuring_2018,avelino_type_2019}. Using the reference High-latitude Time Domain Survey (HLTDS) from \citet{rose_reference_2021}, Figure \ref{fig:roman_obs} shows the cumulative fraction of all \textit{Roman} HLTDS observations usable with SALT3 and SALT3-NIR. At $z\gtrsim1$ the models are equivalent, as the \textit{Roman} filters probe the rest-frame optical, but at $z\lesssim0.5$ in particular SALT3-NIR will be able to fit $\sim20\%$ more data than SALT3, increasing to $\gtrsim85\%$ at $z\lesssim0.1$. Harnessing these lower-$z$ \SNeIa will be particularly valuable for constraints on time-varying dark energy, as they have greater constraining power than other probes at these redshifts \citep[e.g.,][]{jassal_wmap_2005,sendra_supernova_2012}.

The \saltshaker training process is flexible, and will easily accommodate new NIR data in the future to continually refine the SALT3-NIR model as more data become available. \textit{Roman} in particular will provide a wealth of high-precision rest-frame NIR data, including hundreds or thousands of spectra \citep{rose_reference_2021}, which will drastically improve the SALT3-NIR training. While a \roman-only NIR training sample for SALT is plausible in the future, the work herein is necessary now as a robust NIR training methodology, to enable more accurate survey optimization, and produce early-mission \SNIa photometric classification. In advance of \roman there are also a number of ground-based NIR surveys with recent or upcoming public data releases that will be added to the training sample \citep[e.g.,][Hsiao et al., Peterson et al., and Rubin et al. in prep.]{johansson_near-infrared_2021}. Training SALT3-NIR in the \saltshaker framework enables rigorous tests of potential systematic uncertainties arising from the training sample itself, which will be critical as new NIR data are created from future observatories (e.g., Dai et al. and Jones et al., in preparation). 

The next decade of SN\,Ia cosmology is expected to bring a new barrage of data in the NIR, which we are now prepared to harness for improved SN\,Ia light-curve standardization. SALT3-NIR currently provides the only SED-based, open-source means of simulating, classifying, and analyzing these new SNe\,Ia in coming years\footnote{\url{https://doi.org/10.5281/zenodo.7068818}}. 

\acknowledgements{This work was completed in part with resources provided by the University of Chicago’s Research Computing Center. M.D. is supported by the Horizon Fellowship at the Johns Hopkins University. D.S. is supported by DOE grant DE-SC0021962 and the David and Lucile Packard Foundation (grant 2019-69659). L.G. acknowledges financial support from the Spanish Ministerio de Ciencia e Innovaci\'on (MCIN), the Agencia Estatal de Investigaci\'on (AEI) 10.13039/501100011033, and the European Social Fund (ESF) ``Investing in your future" under the 2019 Ram\'on y Cajal program RYC2019-027683-I and the PID2020-115253GA-I00 HOSTFLOWS project, from Centro Superior de Investigaciones Cient\'ificas (CSIC) under the PIE project 20215AT016, and the program Unidad de Excelencia Mar\'ia de Maeztu CEX2020-001058-M. Support for D.O.J. was provided through NASA Hubble Fellowship grant HF2-51462.001, awarded by the Space Telescope Science Institute (STScI), which is operated by the Association of Universities for Research in Astronomy, Inc., under NASA contract NAS5-26555. A.V.F.'s team at UC Berkeley received financial assistance from the Miller Institute for Basic Research in Science (where A.V.F. was a Miller Senior Fellow), the Christopher R. Redlich Fund, and numerous individual donors. The UC Santa Cruz team is supported in part by NASA grants 14-WPS14-0048, NNG16PJ34C, and NNG17PX03C, NSF grants AST-1518052 and AST-1815935, the Gordon and Betty Moore Foundation, the Heising-Simons Foundation, and by a fellowship from the David and Lucile Packard Foundation to R.J.F. This paper is based in part on observations with the NASA/ESA {\it Hubble Space Telescope} obtained from the Mikulski Archive for Space Telescopes at STScI; 
support was provided through programs {\it HST}-GO-15889 and GO-16234 (PI S. Jha), {\it HST}-GO-16706 (PI S. Gezari), and {\it HST}-AR-15808.
This work was supported in part by NASA Keck Data Awards 2020B\_N141 and 2021A\_N147 (PI S. Jha), administered by the NASA Exoplanet Science Institute. The NIRES data used herein were obtained at the W. M. Keck Observatory from telescope time allocated to NASA through the agency's scientific partnership with the California Institute of Technology and the University of California; the Observatory was made possible by the generous financial support of the W. M. Keck Foundation. D.A.H. is supported by NSF grant 1911225. J.B. is supported by NSF grants AST-1911151 and AST-1911225, as well as by NASA grant 80NSSC19kf1639.
 I.P.-F. acknowledges support from the Spanish State Research Agency (AEI) under grant number ESP2017-86582-C4-2-R. F.P. acknowledges support from the Spanish State Research Agency (AEI) under grant number PID2019-105552RB-C43.}

\clearpage

\bibliographystyle{aas}

\begin{thebibliography}{}
\expandafter\ifx\csname natexlab\endcsname\relax\def\natexlab#1{#1}\fi
\providecommand{\url}[1]{\href{#1}{#1}}
\providecommand{\dodoi}[1]{doi:~\href{http://doi.org/#1}{\nolinkurl{#1}}}
\providecommand{\doeprint}[1]{\href{http://ascl.net/#1}{\nolinkurl{http://ascl.net/#1}}}
\providecommand{\doarXiv}[1]{\href{https://arxiv.org/abs/#1}{\nolinkurl{https://arxiv.org/abs/#1}}}
\providecommand{\apj}{ApJ}

\bibitem[{Abbott {et~al.}(2019)Abbott, Allam, Andersen, Angus, Asorey, Avelino,
  Avila, Bassett, Bechtol, Bernstein, Bertin, Brooks, Brout, Brown, Burke,
  Calcino, Rosell, Carollo, Carrasco~Kind, Carretero, Casas, Castander,
  Cawthon, Challis, Childress, Clocchiatti, Cunha, D’Andrea, da~Costa, Davis,
  Davis, De~Vicente, DePoy, Desai, Diehl, Doel, Drlica-Wagner, Eifler, Evrard,
  Fernandez, Filippenko, Finley, Flaugher, Foley, Fosalba, Frieman, Galbany,
  García-Bellido, Gaztanaga, Giannantonio, Glazebrook, Goldstein,
  González-Gaitán, Gruen, Gruendl, Gschwend, Gupta, Gutierrez, Hartley,
  Hinton, Hollowood, Honscheid, Hoormann, Hoyle, James, Jeltema, Johnson,
  Johnson, Kasai, Kent, Kessler, Kim, Kirshner, Kovacs, Krause, Kron, Kuehn,
  Kuhlmann, Kuropatkin, Lahav, Lasker, Lewis, Li, Lidman, Lima, Lin, Macaulay,
  Maia, Mandel, March, Marriner, Marshall, Martini, Menanteau, Miller, Miquel,
  Miranda, Mohr, Morganson, Muthukrishna, Möller, Neilsen, Nichol, Nord,
  Nugent, Ogando, Palmese, Pan, Plazas, Pursiainen, Romer, Roodman, Rozo,
  Rykoff, Sako, Sanchez, Scarpine, Schindler, Schubnell, Scolnic, Serrano,
  Sevilla-Noarbe, Sharp, Smith, Soares-Santos, Sobreira, Sommer, Spinka,
  Suchyta, Sullivan, Swann, Tarle, Thomas, Thomas, Troxel, Tucker, Uddin,
  Walker, Wester, Wiseman, Wolf, Yanny, Zhang, Zhang, \& {DES
  Collaboration}}]{abbott_first_2019}
Abbott, T. M.~C., Allam, S., Andersen, P., {et~al.} 2019, ApJ, 872, L30, \dodoi{10.3847/2041-8213/ab04fa}

\bibitem[{Astier {et~al.}(2006)Astier, Guy, Regnault, Pain, Aubourg, Balam,
  Basa, Carlberg, Fabbro, Fouchez, Hook, Howell, Lafoux, Neill,
  Palanque-Delabrouille, Perrett, Pritchet, Rich, Sullivan, Taillet, Aldering,
  Antilogus, Arsenijevic, Balland, Baumont, Bronder, Courtois, Ellis, Filiol,
  Gonçalves, Goobar, Guide, Hardin, Lusset, Lidman, McMahon, Mouchet, Mourao,
  Perlmutter, Ripoche, Tao, \& Walton}]{astier_supernova_2006}
Astier, P., Guy, J., Regnault, N., {et~al.} 2006, A\&A,
  447, 31, \dodoi{10.1051/0004-6361:20054185}

\bibitem[{Avelino {et~al.}(2019)Avelino, Friedman, Mandel, Jones, Challis, \&
  Kirshner}]{avelino_type_2019}
Avelino, A., Friedman, A.~S., Mandel, K.~S., {et~al.} 2019, ApJ, 887, 106, \dodoi{10.3847/1538-4357/ab2a16}

\bibitem[{Betoule {et~al.}(2014)Betoule, Kessler, Guy, Mosher, Hardin, Biswas,
  Astier, El-Hage, Konig, Kuhlmann, Marriner, Pain, Regnault, Balland, Bassett,
  Brown, Campbell, Carlberg, Cellier-Holzem, Cinabro, Conley, D’Andrea,
  DePoy, Doi, Ellis, Fabbro, Filippenko, Foley, Frieman, Fouchez, Galbany,
  Goobar, Gupta, Hill, Hlozek, Hogan, Hook, Howell, Jha, Le~Guillou, Leloudas,
  Lidman, Marshall, Möller, Mourão, Neveu, Nichol, Olmstead,
  Palanque-Delabrouille, Perlmutter, Prieto, Pritchet, Richmond, Riess,
  Ruhlmann-Kleider, Sako, Schahmaneche, Schneider, Smith, Sollerman, Sullivan,
  Walton, \& Wheeler}]{betoule_improved_2014}
Betoule, M., Kessler, R., Guy, J., {et~al.} 2014, A\&A,
  568, A22, \dodoi{10.1051/0004-6361/201423413}

\bibitem[{Brout {et~al.}(2019)Brout, Scolnic, Kessler, D'Andrea, Davis, Gupta,
  Hinton, Kim, Lasker, Lidman, Macaulay, Möller, Nichol, Sako, Smith,
  Sullivan, Zhang, Andersen, Asorey, Avelino, Bassett, Brown, Calcino, Carollo,
  Challis, Childress, Clocchiatti, Filippenko, Foley, Galbany, Glazebrook,
  Hoormann, Kasai, Kirshner, Kuehn, Kuhlmann, Lewis, Mandel, March, Miranda,
  Morganson, Muthukrishna, Nugent, Palmese, Pan, Sharp, Sommer, Swann, Thomas,
  Tucker, Uddin, Wester, Abbott, Allam, Annis, Avila, Bechtol, Bernstein,
  Bertin, Brooks, Burke, Carnero~Rosell, Carrasco~Kind, Carretero, Castander,
  Cunha, da~Costa, Davis, De~Vicente, DePoy, Desai, Diehl, Doel, Drlica-Wagner,
  Eifler, Estrada, Fernandez, Flaugher, Fosalba, Frieman, García-Bellido,
  Gruen, Gruendl, Gutierrez, Hartley, Hollowood, Honscheid, Hoyle, James,
  Jarvis, Jeltema, Krause, Lahav, Li, Lima, Maia, Marriner, Marshall, Martini,
  Menanteau, Miller, Miquel, Ogando, Plazas, Romer, Roodman, Rykoff, Sanchez,
  Santiago, Scarpine, Schubnell, Serrano, Sevilla-Noarbe, Smith, Soares-Santos,
  Sobreira, Suchyta, Swanson, Tarle, Thomas, Troxel, Tucker, Vikram, Walker,
  Zhang, \& {DES Collaboration}}]{brout_first_2019}
Brout, D., Scolnic, D., Kessler, R., {et~al.} 2019, \apj, 874,
  150, \dodoi{10.3847/1538-4357/ab08a0}

\bibitem[{Brout {et~al.}(2021)Brout, Taylor, Scolnic, Wood, Rose, Vincenzi,
  Dwomoh, Lidman, Riess, Ali, Qu, Dai, \& Stubbs}]{brout_pantheon_2021}
Brout, D., Taylor, G., Scolnic, D., {et~al.} 2021, arXiv e-prints,
  arXiv:2112.03864

\bibitem[{Brout {et~al.}(2022)Brout, Scolnic, Popovic, Riess, Zuntz, Kessler,
  Carr, Davis, Hinton, Jones, Kenworthy, Peterson, Said, Taylor, Ali,
  Armstrong, Charvu, Dwomoh, Palmese, Qu, Rose, Stubbs, Vincenzi, Wood, Brown,
  Chen, Chambers, Coulter, Dai, Dimitriadis, Filippenko, Foley, Jha, Kelsey,
  Kirshner, Möller, Muir, Nadathur, Pan, Rest, Rojas-Bravo, Sako, Siebert,
  Smith, Stahl, \& Wiseman}]{brout_pantheon_2022}
Brout, D., Scolnic, D., Popovic, B., {et~al.} 2022, arXiv e-prints,
  arXiv:2202.04077

\bibitem[{Burns {et~al.}(2011)Burns, Stritzinger, Phillips, Kattner, Persson,
  Madore, Freedman, Boldt, Campillay, Contreras, Folatelli, Gonzalez,
  Krzeminski, Morrell, Salgado, \& Suntzeff}]{burns_carnegie_2011}
Burns, C.~R., Stritzinger, M., Phillips, M.~M., {et~al.} 2011, The Astronomical
  Journal, 141, 19, \dodoi{10.1088/0004-6256/141/1/19}

\bibitem[{Burns {et~al.}(2014)Burns, Stritzinger, Phillips, Hsiao, Contreras,
  Persson, Folatelli, Boldt, Campillay, Castellón, Freedman, Madore, Morrell,
  Salgado, \& Suntzeff}]{burns_carnegie_2014}
---. 2014, ApJ, 789, 32,
  \dodoi{10.1088/0004-637X/789/1/32}

\bibitem[{Conley {et~al.}(2011)Conley, Guy, Sullivan, Regnault, Astier,
  Balland, Basa, Carlberg, Fouchez, Hardin, Hook, Howell, Pain,
  Palanque-Delabrouille, Perrett, Pritchet, Rich, Ruhlmann-Kleider, Balam,
  Baumont, Ellis, Fabbro, Fakhouri, Fourmanoit, González-Gaitán, Graham,
  Hudson, Hsiao, Kronborg, Lidman, Mourao, Neill, Perlmutter, Ripoche, Suzuki,
  \& Walker}]{conley_supernova_2011}
Conley, A., Guy, J., Sullivan, M., {et~al.} 2011, ApJ
  Supplement Series, 192, 1, \dodoi{10.1088/0067-0049/192/1/1}

\bibitem[{Contreras {et~al.}(2010)Contreras, Hamuy, Phillips, Folatelli,
  Suntzeff, Persson, Stritzinger, Boldt, González, Krzeminski, Morrell, Roth,
  Salgado, Maureira, Burns, Freedman, Madore, Murphy, Wyatt, Li, \&
  Filippenko}]{contreras_carnegie_2010}
Contreras, C., Hamuy, M., Phillips, M.~M., {et~al.} 2010, \aj,
  139, 519, \dodoi{10.1088/0004-6256/139/2/519}

\bibitem[{Currie {et~al.}(2020)Currie, Rubin, Aldering, Deustua, Fruchter, \&
  Perlmutter}]{currie_evaluating_2020}
Currie, M., Rubin, D., Aldering, G., {et~al.} 2020, arXiv e-prints,
  arXiv:2007.02458

\bibitem[{Dettman {et~al.}(2021)Dettman, Jha, Dai, Foley, Rest, Scolnic,
  Siebert, Chambers, Coulter, Huber, Johnson, Jones, Kilpatrick, Kirshner, Pan,
  Riess, \& Schultz}]{dettman_foundation_2021}
Dettman, K.~G., Jha, S.~W., Dai, M., {et~al.} 2021, arXiv:2102.06524

\bibitem[{Dhawan {et~al.}(2018)Dhawan, Jha, \&
  Leibundgut}]{dhawan_measuring_2018}
Dhawan, S., Jha, S.~W., \& Leibundgut, B. 2018, \aap, 609, A72,
  \dodoi{10.1051/0004-6361/201731501}

\bibitem[{Elias {et~al.}(1986)Elias, Matthews, Neugebauer, \&
  Soifer}]{elias_observation_1986}
Elias, J.~H., Matthews, K., Neugebauer, G., \& Soifer, B.~T. 1986, Bulletin of
  the American Astronomical Society, 18, 1016

\bibitem[{Fitzpatrick(1999)}]{fitzpatrick_correcting_1999}
Fitzpatrick, E. 1999, PASP,
  111, 63, \dodoi{10.1086/316293}

\bibitem[{Foley {et~al.}(2018)Foley, Scolnic, Rest, Jha, Pan, Riess, Challis,
  Chambers, Coulter, Dettman, Foley, Fox, Huber, Jones, Kilpatrick, Kirshner,
  Schultz, Siebert, Flewelling, Gibson, Magnier, Miller, Primak, Smartt, Smith,
  Wainscoat, Waters, \& Willman}]{foley_foundation_2018}
Foley, R.~J., Scolnic, D., Rest, A., {et~al.} 2018, MNRAS, 475, 193, \dodoi{10.1093/mnras/stx3136}

\bibitem[{Friedman {et~al.}(2015)Friedman, Wood-Vasey, Marion, Challis, Mandel,
  Bloom, Modjaz, Narayan, Hicken, Foley, Klein, Starr, Morgan, Rest, Blake,
  Miller, Falco, Wyatt, Mink, Skrutskie, \& Kirshner}]{friedman_cfair2_2015}
Friedman, A.~S., Wood-Vasey, W.~M., Marion, G.~H., {et~al.} 2015,
  \apjs, 220, 9, \dodoi{10.1088/0067-0049/220/1/9}

\bibitem[{Garnavich {et~al.}(1998)Garnavich, Jha, Challis, Clocchiatti,
  Diercks, Filippenko, Gilliland, Hogan, Kirshner, Leibundgut, Phillips, Reiss,
  Riess, Schmidt, Schommer, Smith, Spyromilio, Stubbs, Suntzeff, Tonry, \&
  Carroll}]{garnavich_supernova_1998}
Garnavich, P.~M., Jha, S., Challis, P., {et~al.} 1998, \apj,
  509, 74, \dodoi{10.1086/306495}

\bibitem[{Guy {et~al.}(2005)Guy, Astier, Nobili, Regnault, \&
  Pain}]{guy_salt:_2005}
Guy, J., Astier, P., Nobili, S., Regnault, N., \& Pain, R. 2005, A\&A, 443, 781, \dodoi{10.1051/0004-6361:20053025}

\bibitem[{Guy {et~al.}(2007)Guy, Astier, Baumont, Hardin, Pain, Regnault, Basa,
  Carlberg, Conley, Fabbro, Fouchez, Hook, Howell, Perrett, Pritchet, Rich,
  Sullivan, Antilogus, Aubourg, Bazin, Bronder, Filiol, Palanque-Delabrouille,
  Ripoche, \& Ruhlmann-Kleider}]{guy_salt2:_2007}
Guy, J., Astier, P., Baumont, S., {et~al.} 2007, A\&A,
  466, 11, \dodoi{10.1051/0004-6361:20066930}

\bibitem[{Guy {et~al.}(2010)Guy, Sullivan, Conley, Regnault, Astier, Balland,
  Basa, Carlberg, Fouchez, Hardin, Hook, Howell, Pain, Palanque-Delabrouille,
  Perrett, Pritchet, Rich, Ruhlmann-Kleider, Balam, Baumont, Ellis, Fabbro,
  Fakhouri, Fourmanoit, González-Gaitán, Graham, Hsiao, Kronborg, Lidman,
  Mourao, Perlmutter, Ripoche, Suzuki, \& Walker}]{guy_supernova_2010}
Guy, J., Sullivan, M., Conley, A., {et~al.} 2010, A\&A,
  523, A7, \dodoi{10.1051/0004-6361/201014468}

\bibitem[{Hamuy {et~al.}(1996)Hamuy, Phillips, Suntzeff, Schommer, Maza, \&
  Aviles}]{hamuy_hubble_1996}
Hamuy, M., Phillips, M.~M., Suntzeff, N.~B., {et~al.} 1996, The Astronomical
  Journal, 112, 2398, \dodoi{10.1086/118191}

\bibitem[{Hicken {et~al.}(2009)Hicken, Wood-Vasey, Blondin, Challis, Jha,
  Kelly, Rest, \& Kirshner}]{hicken_improved_2009}
Hicken, M., Wood-Vasey, W.~M., Blondin, S., {et~al.} 2009, \apj,
  700, 1097, \dodoi{10.1088/0004-637X/700/2/1097}

\bibitem[{Hicken {et~al.}(2012)Hicken, Challis, Kirshner, Rest, Cramer,
  Wood-Vasey, Bakos, Berlind, Brown, Caldwell, Calkins, Currie, de~Kleer,
  Esquerdo, Everett, Falco, Fernandez, Friedman, Groner, Hartman, Holman,
  Hutchins, Keys, Kipping, Latham, Marion, Narayan, Pahre, Pal, Peters,
  Perumpilly, Ripman, Sipocz, Szentgyorgyi, Tang, Torres, Vaz, Wolk, \&
  Zezas}]{hicken_cfa4_2012}
Hicken, M., Challis, P., Kirshner, R.~P., {et~al.} 2012, \apjs,
  200, 12, \dodoi{10.1088/0067-0049/200/2/12}

\bibitem[{Hodgkin {et~al.}(2009)Hodgkin, Irwin, Hewett, \&
  Warren}]{hodgkin_ukirt_2009}
Hodgkin, S.~T., Irwin, M.~J., Hewett, P.~C., \& Warren, S.~J. 2009,
  \mnras, 394, 675, \dodoi{10.1111/j.1365-2966.2008.14387.x}

\bibitem[{Holtzman {et~al.}(2008)Holtzman, Marriner, Kessler, Sako, Dilday,
  Frieman, Schneider, Bassett, Becker, Cinabro, DeJongh, Depoy, Doi, Garnavich,
  Hogan, Jha, Konishi, Lampeitl, Marshall, McGinnis, Miknaitis, Nichol, Prieto,
  Riess, Richmond, Romani, Smith, Takanashi, Tokita, van~der Heyden, Yasuda, \&
  Zheng}]{holtzman_sloan_2008}
Holtzman, J.~A., Marriner, J., Kessler, R., {et~al.} 2008, The Astronomical
  Journal, 136, 2306, \dodoi{10.1088/0004-6256/136/6/2306}

\bibitem[{Hosseinzadeh {et~al.}(2020)Hosseinzadeh, Dauphin, Villar, Berger,
  Jones, Challis, Chornock, Drout, Foley, Kirshner, Lunnan, Margutti,
  Milisavljevic, Pan, Rest, Scolnic, Magnier, Metcalfe, Wainscoat, \&
  Waters}]{hosseinzadeh_photometric_2020}
Hosseinzadeh, G., Dauphin, F., Villar, V.~A., {et~al.} 2020,
  \apj, 905, 93, \dodoi{10.3847/1538-4357/abc42b}

\bibitem[{Hounsell {et~al.}(2018)Hounsell, Scolnic, Foley, Kessler, Miranda,
  Avelino, Bohlin, Filippenko, Frieman, Jha, Kelly, Kirshner, Mandel, Rest,
  Riess, Rodney, \& Strolger}]{hounsell_simulations_2018}
Hounsell, R., Scolnic, D., Foley, R.~J., {et~al.} 2018, ApJ, 867, 23, \dodoi{10.3847/1538-4357/aac08b}

\bibitem[{Hsiao {et~al.}(2007)Hsiao, Conley, Howell, Sullivan, Pritchet,
  Carlberg, Nugent, \& Phillips}]{hsiao_k_2007}
Hsiao, E.~Y., Conley, A., Howell, D.~A., {et~al.} 2007, ApJ, 663, 1187, \dodoi{10.1086/518232}

\bibitem[{Ivanov {et~al.}(2000)Ivanov, Hamuy, \& Pinto}]{ivanov_relation_2000}
Ivanov, V.~D., Hamuy, M., \& Pinto, P.~A. 2000, \apj, 542, 588,
  \dodoi{10.1086/317060}

\bibitem[{Jassal {et~al.}(2005)Jassal, Bagla, \&
  Padmanabhan}]{jassal_wmap_2005}
Jassal, H.~K., Bagla, J.~S., \& Padmanabhan, T. 2005, \mnras,
  356, L11, \dodoi{10.1111/j.1745-3933.2005.08577.x}

\bibitem[{Jha {et~al.}(2007)Jha, Riess, \& Kirshner}]{jha_improved_2007}
Jha, S., Riess, A.~G., \& Kirshner, R.~P. 2007, ApJ, 659,
  122, \dodoi{10.1086/512054}

\bibitem[{Jha {et~al.}(2006)Jha, Kirshner, Challis, Garnavich, Matheson,
  Soderberg, Graves, Hicken, Alves, Arce, Balog, Barmby, Barton, Berlind,
  Bragg, Briceño, Brown, Buckley, Caldwell, Calkins, Carter, Concannon,
  Donnelly, Eriksen, Fabricant, Falco, Fiore, Garcia, Gómez, Grogin, Groner,
  Groot, Haisch, Hartmann, Hergenrother, Holman, Huchra, Jayawardhana, Jerius,
  Kannappan, Kim, Kleyna, Kochanek, Koranyi, Krockenberger, Lada, Luhman, Luu,
  Macri, Mader, Mahdavi, Marengo, Marsden, McLeod, McNamara, Megeath, Moraru,
  Mossman, Muench, Muñoz, Muzerolle, Naranjo, Nelson-Patel, Pahre, Patten,
  Peters, Peters, Raymond, Rines, Schild, Sobczak, Spahr, Stauffer, Stefanik,
  Szentgyorgyi, Tollestrup, Väisänen, Vikhlinin, Wang, Willner, Wolk, Zajac,
  Zhao, \& Stanek}]{jha_ubvri_2006}
Jha, S., Kirshner, R.~P., Challis, P., {et~al.} 2006, \aj, 131,
  527, \dodoi{10.1086/497989}

\bibitem[{Johansson {et~al.}(2021)Johansson, Cenko, Fox, Dhawan, Goobar,
  Stanishev, Butler, Lee, Watson, Fremling, Kasliwal, Nugent, Petrushevska,
  Sollerman, Yan, Burke, Hosseinzadeh, Howell, McCully, \&
  Valenti}]{johansson_near-infrared_2021}
Johansson, J., Cenko, S.~B., Fox, O.~D., {et~al.} 2021, \apj,
  923, 237, \dodoi{10.3847/1538-4357/ac2f9e}

\bibitem[{Jones {et~al.}(2017)Jones, Scolnic, Riess, Kessler, Rest, Kirshner,
  Berger, Ortega, Foley, Chornock, Challis, Burgett, Chambers, Draper,
  Flewelling, Huber, Kaiser, Kudritzki, Metcalfe, Wainscoat, \&
  Waters}]{jones_measuring_2017}
Jones, D.~O., Scolnic, D.~M., Riess, A.~G., {et~al.} 2017, \apj,
  843, 6, \dodoi{10.3847/1538-4357/aa767b}

\bibitem[{Jones {et~al.}(2019)Jones, Scolnic, Foley, Rest, Kessler, Challis,
  Chambers, Coulter, Dettman, Foley, Huber, Jha, Johnson, Kilpatrick, Kirshner,
  Manuel, Narayan, Pan, Riess, Schultz, Siebert, Berger, Chornock, Flewelling,
  Magnier, Smartt, Smith, Wainscoat, Waters, \&
  Willman}]{jones_foundation_2019}
Jones, D.~O., Scolnic, D.~M., Foley, R.~J., {et~al.} 2019, ApJ, 881, 19, \dodoi{10.3847/1538-4357/ab2bec}

\bibitem[{Jones {et~al.}(2022)Jones, Mandel, Kirshner, Thorp, Challis, Avelino,
  Brout, Burns, Foley, Pan, Scolnic, Siebert, Chornock, Freedman, Friedman,
  Frieman, Galbany, Hsiao, Kelsey, Marion, Nichol, Nugent, Phillips, Rest,
  Riess, Sako, Smith, Wiseman, \& Wood-Vasey}]{jones_cosmological_2022}
Jones, D.~O., Mandel, K.~S., Kirshner, R.~P., {et~al.} 2022, arXiv e-prints,
  arXiv:2201.07801

\bibitem[{Kelly {et~al.}(2010)Kelly, Hicken, Burke, Mandel, \&
  Kirshner}]{kelly_hubble_2010}
Kelly, P.~L., Hicken, M., Burke, D.~L., Mandel, K.~S., \& Kirshner, R.~P. 2010,
  ApJ, 715, 743, \dodoi{10.1088/0004-637X/715/2/743}

\bibitem[{Kenworthy {et~al.}(2021)Kenworthy, Jones, Dai, Kessler, Scolnic,
  Brout, Siebert, Pierel, Dettman, Dimitriadis, Foley, Jha, Pan, Riess, Rodney,
  \& Rojas-Bravo}]{kenworthy_salt3_2021}
Kenworthy, W.~D., Jones, D.~O., Dai, M., {et~al.} 2021, arXiv e-prints,
  arXiv:2104.07795

\bibitem[{Kessler \& Scolnic(2017)}]{kessler_correcting_2017}
Kessler, R., \& Scolnic, D. 2017, ApJ, 836, 56,
  \dodoi{10.3847/1538-4357/836/1/56}

\bibitem[{Kessler {et~al.}(2009)Kessler, Bernstein, Cinabro, Dilday, Frieman,
  Jha, Kuhlmann, Miknaitis, Sako, Taylor, \& Vanderplas}]{kessler_snana:_2009}
Kessler, R., Bernstein, J.~P., Cinabro, D., {et~al.} 2009, PASP, 121, 1028, \dodoi{10.1086/605984}

\bibitem[{Kessler {et~al.}(2019)Kessler, Brout, D’Andrea, Davis, Hinton, Kim,
  Lasker, Lidman, Macaulay, Möller, Sako, Scolnic, Smith, Sullivan, Zhang,
  Andersen, Asorey, Avelino, Calcino, Carollo, Challis, Childress, Clocchiatti,
  Crawford, Filippenko, Foley, Glazebrook, Hoormann, Kasai, Kirshner, Lewis,
  Mandel, March, Morganson, Muthukrishna, Nugent, Pan, Sommer, Swann, Thomas,
  Tucker, Uddin, Abbott, Allam, Annis, Avila, Banerji, Bechtol, Bertin, Brooks,
  Buckley-Geer, Burke, Carnero Rosell, Carrasco Kind, Carretero, Castander,
  Crocce, da Costa, Davis, De Vicente, Desai, Diehl, Doel, Eifler, Flaugher,
  Fosalba, Frieman, García-Bellido, Gaztanaga, Gerdes, Gruen, Gruendl,
  Gutierrez, Hartley, Hollowood, Honscheid, James, Johnson, Johnson, Krause,
  Kuehn, Kuropatkin, Lahav, Li, Lima, Marshall, Martini, Menanteau, Miller,
  Miquel, Nord, Plazas, Roodman, Sanchez, Scarpine, Schindler, Schubnell,
  Serrano, Sevilla-Noarbe, Soares-Santos, Sobreira, Suchyta, Tarle, Thomas,
  Walker, Zhang, \& {DES Collaboration}}]{kessler_first_2019}
Kessler, R., Brout, D., D’Andrea, C.~B., {et~al.} 2019, MNRAS, 485, 1171, \dodoi{10.1093/mnras/stz463}

\bibitem[{Konchady {et~al.}(2022)Konchady, Oelkers, Jones, Yuan, Macri,
  Peterson, \& Riess}]{konchady_h-band_2022}
Konchady, T., Oelkers, R.~J., Jones, D.~O., {et~al.} 2022,
  \apjs, 258, 24, \dodoi{10.3847/1538-4365/ac41d3}

\bibitem[{Krisciunas {et~al.}(2004{\natexlab{a}})Krisciunas, Phillips, \&
  Suntzeff}]{krisciunas_hubble_2004}
Krisciunas, K., Phillips, M.~M., \& Suntzeff, N.~B. 2004{\natexlab{a}},
  \apjl, 602, L81, \dodoi{10.1086/382731}

\bibitem[{Krisciunas {et~al.}(2003)Krisciunas, Suntzeff, Candia, Arenas,
  Espinoza, Gonzalez, Gonzalez, Höflich, Landolt, Phillips, \&
  Pizarro}]{krisciunas_optical_2003}
Krisciunas, K., Suntzeff, N.~B., Candia, P., {et~al.} 2003, \aj,
  125, 166, \dodoi{10.1086/345571}

\bibitem[{Krisciunas {et~al.}(2004{\natexlab{b}})Krisciunas, Suntzeff,
  Phillips, Candia, Prieto, Antezana, Chassagne, Chen, Dickinson, Eisenhardt,
  Espinoza, Garnavich, González, Harrison, Hamuy, Ivanov, Krzemiński, Kulesa,
  McCarthy, Moro-Martín, Muena, Noriega-Crespo, Persson, Pinto, Roth,
  Rubenstein, Stanford, Stringfellow, Zapata, Porter, \&
  Wischnjewsky}]{krisciunas_optical_2004}
Krisciunas, K., Suntzeff, N.~B., Phillips, M.~M., {et~al.} 2004{\natexlab{b}},
  \aj, 128, 3034, \dodoi{10.1086/425629}

\bibitem[{Krisciunas {et~al.}(2007)Krisciunas, Garnavich, Stanishev, Suntzeff,
  Prieto, Espinoza, Gonzalez, Salvo, Elias de~la Rosa, Smartt, Maund, \&
  Kudritzki}]{krisciunas_type_2007}
Krisciunas, K., Garnavich, P.~M., Stanishev, V., {et~al.} 2007,
  \aj, 133, 58, \dodoi{10.1086/509126}

\bibitem[{Krisciunas {et~al.}(2017)Krisciunas, Contreras, Burns, Phillips,
  Stritzinger, Morrell, Hamuy, Anais, Boldt, Busta, Campillay, Castellón,
  Folatelli, Freedman, González, Hsiao, Krzeminski, Persson, Roth, Salgado,
  Serón, Suntzeff, Torres, Filippenko, Li, Madore, DePoy, Marshall, Rheault,
  \& Villanueva}]{krisciunas_carnegie_2017}
Krisciunas, K., Contreras, C., Burns, C.~R., {et~al.} 2017, \aj,
  154, 211, \dodoi{10.3847/1538-3881/aa8df0}

\bibitem[{Leget {et~al.}(2020)Leget, Gangler, Mondon, Aldering, Antilogus,
  Aragon, Bailey, Baltay, Barbary, Bongard, Boone, Buton, Chotard, Copin,
  Dixon, Fagrelius, Feindt, Fouchez, Hayden, Hillebrandt, Kim, Kowalski,
  Kuesters, Lombardo, Lin, Nordin, Pain, Pecontal, Pereira, Perlmutter, Ponder,
  Pruzhinskaya, Rabinowitz, Rigault, Runge, Rubin, Saunders, Says, Smadja,
  Sofiatti, Suzuki, Taubenberger, Tao, \& Thomas}]{leget_sugar_2020}
Leget, P.-F., Gangler, E., Mondon, F., {et~al.} 2020, A\&A, 636, A46, \dodoi{10.1051/0004-6361/201834954}

\bibitem[{Leloudas {et~al.}(2009)Leloudas, Stritzinger, Sollerman, Burns,
  Kozma, Krisciunas, Maund, Milne, Filippenko, Fransson, Ganeshalingam, Hamuy,
  Li, Phillips, Schmidt, Skottfelt, Taubenberger, Boldt, Fynbo, Gonzalez,
  Salvo, \& Thomas-Osip}]{leloudas_normal_2009}
Leloudas, G., Stritzinger, M.~D., Sollerman, J., {et~al.} 2009,
  \aap, 505, 265, \dodoi{10.1051/0004-6361/200912364}

\bibitem[{Mandel {et~al.}(2011)Mandel, Narayan, \& Kirshner}]{mandel_type_2011}
Mandel, K.~S., Narayan, G., \& Kirshner, R.~P. 2011, \apj, 731,
  120, \dodoi{10.1088/0004-637X/731/2/120}

\bibitem[{Mandel {et~al.}(2020)Mandel, Thorp, Narayan, Friedman, \&
  Avelino}]{mandel_hierarchical_2020}
Mandel, K.~S., Thorp, S., Narayan, G., Friedman, A.~S., \& Avelino, A. 2020,
  arXiv e-prints, arXiv:2008.07538

\bibitem[{Mandel {et~al.}(2022)Mandel, Thorp, Narayan, Friedman, \&
  Avelino}]{mandel_hierarchical_2022}
---. 2022, \mnras, 510, 3939, \dodoi{10.1093/mnras/stab3496}

\bibitem[{Marion {et~al.}(2016)Marion, Brown, Vinkó, Silverman, Sand, Challis,
  Kirshner, Wheeler, Berlind, Brown, Calkins, Camacho, Dhungana, Foley,
  Friedman, Graham, Howell, Hsiao, Irwin, Jha, Kehoe, Macri, Maeda, Mandel,
  McCully, Pandya, Rines, Wilhelmy, \& Zheng}]{marion_sn_2016}
Marion, G.~H., Brown, P.~J., Vinkó, J., {et~al.} 2016, \apj,
  820, 92, \dodoi{10.3847/0004-637X/820/2/92}

\bibitem[{Marriner {et~al.}(2011)Marriner, Bernstein, Kessler, Lampeitl,
  Miquel, Mosher, Nichol, Sako, Schneider, \& Smith}]{marriner_more_2011}
Marriner, J., Bernstein, J.~P., Kessler, R., {et~al.} 2011,
  \apj, 740, 72, \dodoi{10.1088/0004-637X/740/2/72}

\bibitem[{Masci {et~al.}(2019)Masci, Laher, Rusholme, Shupe, Groom, Surace,
  Jackson, Monkewitz, Beck, Flynn, Terek, Landry, Hacopians, Desai, Howell,
  Brooke, Imel, Wachter, Ye, Lin, Cenko, Cunningham, Rebbapragada, Bue, Miller,
  Mahabal, Bellm, Patterson, Jurić, Golkhou, Ofek, Walters, Graham, Kasliwal,
  Dekany, Kupfer, Burdge, Cannella, Barlow, Van~Sistine, Giomi, Fremling,
  Blagorodnova, Levitan, Riddle, Smith, Helou, Prince, \&
  Kulkarni}]{masci_zwicky_2019}
Masci, F.~J., Laher, R.~R., Rusholme, B., {et~al.} 2019, \pasp,
  131, 018003, \dodoi{10.1088/1538-3873/aae8ac}

\bibitem[{Mosher {et~al.}(2014)Mosher, Guy, Kessler, Astier, Marriner, Betoule,
  Sako, El-Hage, Biswas, Pain, Kuhlmann, Regnault, Frieman, \&
  Schneider}]{mosher_cosmological_2014}
Mosher, J., Guy, J., Kessler, R., {et~al.} 2014, ApJ,
  793, 16, \dodoi{10.1088/0004-637X/793/1/16}

\bibitem[{Perlmutter {et~al.}(1999)Perlmutter, Aldering, Goldhaber, Knop,
  Nugent, Castro, Deustua, Fabbro, Goobar, Groom, Hook, Kim, Kim, Lee, Nunes,
  Pain, Pennypacker, Quimby, Lidman, Ellis, Irwin, McMahon, Ruiz-Lapuente,
  Walton, Schaefer, Boyle, Filippenko, Matheson, Fruchter, Panagia, Newberg,
  Couch, \& Project}]{perlmutter_measurements_1999}
Perlmutter, S., Aldering, G., Goldhaber, G., {et~al.} 1999, ApJ, 517, 565, \dodoi{10.1086/307221}

\bibitem[{Pierel {et~al.}(2018)Pierel, Rodney, Avelino, Bianco, Filippenko,
  Foley, Friedman, Hicken, Hounsell, Jha, Kessler, Kirshner, Mandel, Narayan,
  Scolnic, \& Strolger}]{pierel_extending_2018}
Pierel, J. D.~R., Rodney, S., Avelino, A., {et~al.} 2018, PASP, 130, 114504,
  \dodoi{10.1088/1538-3873/aadb7a}

\bibitem[{Pignata {et~al.}(2008)Pignata, Benetti, Mazzali, Kotak, Patat,
  Meikle, Stehle, Leibundgut, Suntzeff, Buson, Cappellaro, Clocchiatti, Hamuy,
  Maza, Mendez, Ruiz-Lapuente, Salvo, Schmidt, Turatto, \&
  Hillebrandt}]{pignata_optical_2008}
Pignata, G., Benetti, S., Mazzali, P.~A., {et~al.} 2008, \mnras,
  388, 971, \dodoi{10.1111/j.1365-2966.2008.13434.x}

\bibitem[{Rest {et~al.}(2005)Rest, Stubbs, Becker, Miknaitis, Miceli,
  Covarrubias, Hawley, Smith, Suntzeff, Olsen, Prieto, Hiriart, Welch, Cook,
  Nikolaev, Huber, Prochtor, Clocchiatti, Minniti, Garg, Challis, Keller, \&
  Schmidt}]{rest_testing_2005}
Rest, A., Stubbs, C., Becker, A.~C., {et~al.} 2005, \apj, 634,
  1103, \dodoi{10.1086/497060}

\bibitem[{Riess {et~al.}(1998)Riess, Filippenko, Challis, Clocchiatti, Diercks,
  Garnavich, Gilliland, Hogan, Jha, Kirshner, Leibundgut, Phillips, Reiss,
  Schmidt, Schommer, Smith, Spyromilio, Stubbs, Suntzeff, \&
  Tonry}]{riess_observational_1998}
Riess, A.~G., Filippenko, A.~V., Challis, P., {et~al.} 1998, The Astronomical
  Journal, 116, 1009, \dodoi{10.1086/300499}

\bibitem[{Riess {et~al.}(1999)Riess, Kirshner, Schmidt, Jha, Challis,
  Garnavich, Esin, Carpenter, Grashius, Schild, Berlind, Huchra, Prosser,
  Falco, Benson, Briceño, Brown, Caldwell, dell'Antonio, Filippenko, Goodman,
  Grogin, Groner, Hughes, Green, Jansen, Kleyna, Luu, Macri, McLeod, McLeod,
  McNamara, McLean, Milone, Mohr, Moraru, Peng, Peters, Prestwich, Stanek,
  Szentgyorgyi, \& Zhao}]{riess_bvri_1999}
Riess, A.~G., Kirshner, R.~P., Schmidt, B.~P., {et~al.} 1999,
  \aj, 117, 707, \dodoi{10.1086/300738}

\bibitem[{Riess {et~al.}(2018)Riess, Rodney, Scolnic, Shafer, Strolger,
  Ferguson, Postman, Graur, Maoz, Jha, Mobasher, Casertano, Hayden, Molino,
  Hjorth, Garnavich, Jones, Kirshner, Koekemoer, Grogin, Brammer, Hemmati,
  Dickinson, Challis, Wolff, Clubb, Filippenko, Nayyeri, U, Koo, Faber,
  Kocevski, Bradley, \& Coe}]{riess_type_2018}
Riess, A.~G., Rodney, S.~A., Scolnic, D.~M., {et~al.} 2018,
  \apj, 853, 126, \dodoi{10.3847/1538-4357/aaa5a9}

\bibitem[{Rose {et~al.}(2021)Rose, Baltay, Hounsell, Macias, Rubin, Scolnic,
  Aldering, Bohlin, Dai, Deustua, Foley, Fruchter, Galbany, Jha, Jones, Joshi,
  Kelly, Kessler, Kirshner, Mandel, Perlmutter, Pierel, Qu, Rabinowitz, Rest,
  Riess, Rodney, Sako, Siebert, Strolger, Suzuki, Thorp, Van~Dyk, Wang, Ward,
  \& Wood-Vasey}]{rose_reference_2021}
Rose, B.~M., Baltay, C., Hounsell, R., {et~al.} 2021, arXiv e-prints,
  arXiv:2111.03081

\bibitem[{Sako {et~al.}(2018)Sako, Bassett, Becker, Brown, Campbell, Wolf,
  Cinabro, D'Andrea, Dawson, DeJongh, Depoy, Dilday, Doi, Filippenko, Fischer,
  Foley, Frieman, Galbany, Garnavich, Goobar, Gupta, Hill, Hayden, Hlozek,
  Holtzman, Hopp, Jha, Kessler, Kollatschny, Leloudas, Marriner, Marshall,
  Miquel, Morokuma, Mosher, Nichol, Nordin, Olmstead, Östman, Prieto,
  Richmond, Romani, Sollerman, Stritzinger, Schneider, Smith, Wheeler, Yasuda,
  \& Zheng}]{sako_data_2018}
Sako, M., Bassett, B., Becker, A.~C., {et~al.} 2018, \pasp, 130,
  064002, \dodoi{10.1088/1538-3873/aab4e0}

\bibitem[{Saunders {et~al.}(2018)Saunders, Aldering, Antilogus, Bailey, Baltay,
  Barbary, Baugh, Boone, Bongard, Buton, Chen, Chotard, Copin, Dixon,
  Fagrelius, Fakhouri, Feindt, Fouchez, Gangler, Hayden, Hillebrandt, Kim,
  Kowalski, Küsters, Leget, Lombardo, Nordin, Pain, Pecontal, Pereira,
  Perlmutter, Rabinowitz, Rigault, Rubin, Runge, Smadja, Sofiatti, Suzuki, Tao,
  Taubenberger, Thomas, Vincenzi, \& {The Nearby Supernova
  Factory}}]{saunders_snemo_2018}
Saunders, C., Aldering, G., Antilogus, P., {et~al.} 2018, ApJ, 869, 167, \dodoi{10.3847/1538-4357/aaec7e}

\bibitem[{Scolnic {et~al.}(2015)Scolnic, Casertano, Riess, Rest, Schlafly,
  Foley, Finkbeiner, Tang, Burgett, Chambers, Draper, Flewelling, Hodapp,
  Huber, Kaiser, Kudritzki, Magnier, Metcalfe, \&
  Stubbs}]{scolnic_supercal_2015}
Scolnic, D., Casertano, S., Riess, A., {et~al.} 2015, \apj, 815,
  117, \dodoi{10.1088/0004-637X/815/2/117}

\bibitem[{Scolnic {et~al.}(2018)Scolnic, Jones, Rest, Pan, Chornock, Foley,
  Huber, Kessler, Narayan, Riess, Rodney, Berger, Brout, Challis, Drout,
  Finkbeiner, Lunnan, Kirshner, Sanders, Schlafly, Smartt, Stubbs, Tonry,
  Wood-Vasey, Foley, Hand, Johnson, Burgett, Chambers, Draper, Hodapp, Kaiser,
  Kudritzki, Magnier, Metcalfe, Bresolin, Gall, Kotak, McCrum, \&
  Smith}]{scolnic_complete_2018}
Scolnic, D.~M., Jones, D.~O., Rest, A., {et~al.} 2018, ApJ, 859, 101, \dodoi{10.3847/1538-4357/aab9bb}

\bibitem[{Sendra \& Lazkoz(2012)}]{sendra_supernova_2012}
Sendra, I., \& Lazkoz, R. 2012, \mnras, 422, 776,
  \dodoi{10.1111/j.1365-2966.2012.20661.x}

\bibitem[{Siebert {et~al.}(2019)Siebert, Foley, Jones, Angulo, Davis, Duarte,
  Strasburger, Conlon, Kazmi, Nishimoto, Schubert, Sun, \&
  Tippens}]{siebert_investigating_2019}
Siebert, M.~R., Foley, R.~J., Jones, D.~O., {et~al.} 2019, MNRAS, 486, 5785, \dodoi{10.1093/mnras/stz1209}

\bibitem[{Silverman {et~al.}(2012)Silverman, Foley, Filippenko, Ganeshalingam,
  Barth, Chornock, Griffith, Kong, Lee, Leonard, Matheson, Miller, Steele,
  Barris, Bloom, Cobb, Coil, Desroches, Gates, Ho, Jha, Kandrashoff, Li,
  Mandel, Modjaz, Moore, Mostardi, Papenkova, Park, Perley, Poznanski, Reuter,
  Scala, Serduke, Shields, Swift, Tonry, Dyk, Wang, \&
  Wong}]{silverman_berkeley_2012}
Silverman, J.~M., Foley, R.~J., Filippenko, A.~V., {et~al.} 2012, MNRAS, 425,
  1789

\bibitem[{Skrutskie {et~al.}(2006)Skrutskie, Cutri, Stiening, Weinberg,
  Schneider, Carpenter, Beichman, Capps, Chester, Elias, Huchra, Liebert,
  Lonsdale, Monet, Price, Seitzer, Jarrett, Kirkpatrick, Gizis, Howard, Evans,
  Fowler, Fullmer, Hurt, Light, Kopan, Marsh, McCallon, Tam, Van~Dyk, \&
  Wheelock}]{skrutskie_two_2006}
Skrutskie, M.~F., Cutri, R.~M., Stiening, R., {et~al.} 2006,
  \aj, 131, 1163, \dodoi{10.1086/498708}

\bibitem[{Stahl {et~al.}(2020)Stahl, Zheng, de~Jaeger, Brink, Filippenko,
  Silverman, Cenko, Clubb, Graham, Halevi, Kelly, Kleiser, Shivvers, Yuk, Cobb,
  Fox, Kandrashoff, Kong, Mauerhan, Wang, \& Wang}]{stahl_berkeley_2020}
Stahl, B.~E., Zheng, W., de~Jaeger, T., {et~al.} 2020, \mnras,
  492, 4325, \dodoi{10.1093/mnras/staa102}

\bibitem[{Stanishev {et~al.}(2007)Stanishev, Goobar, Benetti, Kotak, Pignata,
  Navasardyan, Mazzali, Amanullah, Garavini, Nobili, Qiu, Elias-Rosa,
  Ruiz-Lapuente, Mendez, Meikle, Patat, Pastorello, Altavilla, Gustafsson,
  Harutyunyan, Iijima, Jakobsson, Kichizhieva, Lundqvist, Mattila, Melinder,
  Pavlenko, Pavlyuk, Sollerman, Tsvetkov, Turatto, \&
  Hillebrandt}]{stanishev_sn_2007}
Stanishev, V., Goobar, A., Benetti, S., {et~al.} 2007, \aap,
  469, 645, \dodoi{10.1051/0004-6361:20066020}

\bibitem[{Stritzinger {et~al.}(2010)Stritzinger, Filippenko, Folatelli, Foley,
  Hamuy, Li, Mazzali, Phillips, \& Pignata}]{stritzinger_multi_2010}
Stritzinger, M., Filippenko, A., Folatelli, G., {et~al.} 2010, Multi-wavelength
  spectroscopic study of young {Type} {Ia} supernovae

\bibitem[{Stritzinger {et~al.}(2011)Stritzinger, Phillips, Boldt, Burns,
  Campillay, Contreras, Gonzalez, Folatelli, Morrell, Krzeminski, Roth,
  Salgado, DePoy, Hamuy, Freedman, Madore, Marshall, Persson, Rheault,
  Suntzeff, Villanueva, Li, \& Filippenko}]{stritzinger_carnegie_2011}
Stritzinger, M.~D., Phillips, M.~M., Boldt, L.~N., {et~al.} 2011,
  \aj, 142, 156, \dodoi{10.1088/0004-6256/142/5/156}

\bibitem[{Sullivan {et~al.}(2010)Sullivan, Conley, Howell, Neill, Astier,
  Balland, Basa, Carlberg, Fouchez, Guy, Hardin, Hook, Pain,
  Palanque-Delabrouille, Perrett, Pritchet, Regnault, Rich, Ruhlmann-Kleider,
  Baumont, Hsiao, Kronborg, Lidman, Perlmutter, \&
  Walker}]{sullivan_dependence_2010}
Sullivan, M., Conley, A., Howell, D.~A., {et~al.} 2010, MNRAS, no, \dodoi{10.1111/j.1365-2966.2010.16731.x}

\bibitem[{Taylor {et~al.}(2021)Taylor, Lidman, Tucker, Brout, Hinton, \&
  Kessler}]{taylor_revised_2021}
Taylor, G., Lidman, C., Tucker, B.~E., {et~al.} 2021, \mnras,
  504, 4111, \dodoi{10.1093/mnras/stab962}

\bibitem[{{The Dark Energy Survey
  Collaboration}(2005)}]{the_dark_energy_survey_collaboration_dark_2005}
{The Dark Energy Survey Collaboration}. 2005, arXiv e-prints, astro

\bibitem[{Tonry {et~al.}(2018)Tonry, Denneau, Heinze, Stalder, Smith, Smartt,
  Stubbs, Weiland, \& Rest}]{tonry_atlas_2018}
Tonry, J.~L., Denneau, L., Heinze, A.~N., {et~al.} 2018, \pasp,
  130, 064505, \dodoi{10.1088/1538-3873/aabadf}

\bibitem[{Tripp(1998)}]{tripp_two-parameter_1998}
Tripp, R. 1998, A\&A, 331, 815

\bibitem[{Valentini {et~al.}(2003)Valentini, Di~Carlo, Massi, Dolci, Arkharov,
  Larionov, Pastorello, Di~Paola, Benetti, Cappellaro, Turatto, Pedichini,
  D'Alessio, Caratti~o Garatti, Li~Causi, Speziali, Danziger, \&
  Tornambé}]{valentini_optical_2003}
Valentini, G., Di~Carlo, E., Massi, F., {et~al.} 2003, \apj,
  595, 779, \dodoi{10.1086/377448}

\bibitem[{Villar {et~al.}(2020)Villar, Hosseinzadeh, Berger, Ntampaka, Jones,
  Challis, Chornock, Drout, Foley, Kirshner, Lunnan, Margutti, Milisavljevic,
  Sanders, Pan, Rest, Scolnic, Magnier, Metcalfe, Wainscoat, \&
  Waters}]{villar_superraenn_2020}
Villar, V.~A., Hosseinzadeh, G., Berger, E., {et~al.} 2020,
  \apj, 905, 94, \dodoi{10.3847/1538-4357/abc6fd}

\bibitem[{Wood-Vasey {et~al.}(2008)Wood-Vasey, Friedman, Bloom, Hicken, Modjaz,
  Kirshner, Starr, Blake, Falco, Szentgyorgyi, Challis, Blondin, Mandel, \&
  Rest}]{woodvasey_type_2008}
Wood-Vasey, W.~M., Friedman, A.~S., Bloom, J.~S., {et~al.} 2008,
  \apj, 689, 377, \dodoi{10.1086/592374}

\end{thebibliography}

\end{document}